\documentclass[twocol]{ametsocV5}

\usepackage{amsmath,amsfonts,amssymb,bm}
\usepackage{mathptmx}
\usepackage{newtxtext}
\usepackage{newtxmath}
\usepackage{amsmath}
\usepackage{grffile}

\bibpunct{(}{)}{;}{a}{}{,}

\title{Revisiting the Quasi Biennial Oscillation as Seen in ERA5.\\
Part I: Description and Momentum Budget}

\authors{Hamid A. Pahlavan\correspondingauthor{Hamid A. Pahlavan, pahlavan@uw.edu}, Qiang Fu, John M. Wallace}

\affiliation{Department of Atmospheric Sciences, University of Washington, Seattle, Washington}

\extraauthor{George N. Kiladis}

\extraaffil{ Earth System Research Laboratory, NOAA, Boulder, Colorado}

\abstract{The dynamics and momentum budget of the quasi-biennial oscillation (QBO) are examined in the ERA5 reanalysis and compared with those in ERA-I. Because of ERA5\textquotesingle s higher spatial resolution it is capable of resolving a broader spectrum of atmospheric waves and allows for a better representation of the wave-mean flow interactions, both of which are of crucial importance for QBO studies. It is shown that the QBO-induced mean meridional circulation, which is mainly confined to the winter hemisphere, is strong enough to interrupt the tropical upwelling during the descent of the westerly shear zones. Since the momentum advection tends to damp the QBO, the wave forcing is responsible for both the downward propagation and for the maintenance of the QBO. It is shown that half the required wave forcing is provided by resolved waves during the descent of both westerly and easterly regimes. Planetary-scale waves account for most of the resolved wave forcing of the descent of westerly shear zones and small-scale gravity (SSG) waves with wavelengths shorter than 2000 km account for the remainder. SSG waves account for most of the resolved forcing of the descent of the easterly shear zones. The representation of the mean fields in the QBO is very similar in ERA5 and ERA-I but the resolved wave forcing is shown to be substantially stronger in ERA5. The contributions of the various equatorially-trapped wave modes to the QBO forcing are documented in Part II.}

\begin{document}

\maketitle

\section{Introduction}

The quasi-biennial oscillation (QBO) is the primary mode of interannual variability in the tropical and subtropical stratosphere (e.g., Baldwin et al. 2001). It is characterized by the downward propagation of successive westerly and easterly wind regimes with an average period of about 28 months. It has come to be recognized as resulting from the interaction between the axisymmetric flow and a broad spectrum of waves dispersing upward from below (Lindzen and Holton 1968; Holton and Lindzen 1972; Dunkerton 1997; Giorgetta et al. 2002). Although the QBO is a tropical stratospheric phenomenon, it influences the circulation and the interannual variability of the global stratosphere and lower mesosphere, and has a discernible effect on the weather and climate in the troposphere. A detailed characterization of the QBO and its impact can be found in the review of Baldwin et al. (2001) and in Anstey et al. (2020).

Since their emergence in the mid- 1990\textquotesingle s, global reanalyses by the world\textquotesingle s leading centers for operational numerical weather prediction have been among the most widely used datasets in the geosciences. Swinbank and O\textquotesingle Neill (1994) and Randel et al. (1999) were able to obtain a dynamically consistent QBO in the analyses produced at the U. K. Met. Office in support of the Upper Atmosphere Research Satellite (UARS) mission. Pawson and Fiorino (1998) investigated the QBO in National Centers for Environmental Prediction-National Center for Atmospheric Research (NCEP-NCAR) reanalyses and European Centre for Medium-Range Weather Forecasts (ECMWF) Reanalysis (ERA-15) and found the axisymmetric zonal wind on the equator to be in good agreement with radiosonde winds at Singapore (1.8\textdegree N, 104.8\textdegree E)  at 30 hPa and below, though the westerly regimes in the reanalysis were generally weaker than observations.

Using singular vector decomposition, Ribera et al. (2004) were able to discern coherent QBO-induced mean meridional circulations in the NCEP-NCAR reanalyses. Pascoe et al. (2005) examined the amplitude, period and vertical structure of the QBO in the ERA-40 data set and were even able to resolve the cycle-to-cycle variability of the QBO seen in the station data. Coy et al. (2016) examined the structure, dynamics, forcing, and ozone signal of the QBO in the NASA Modern-Era Retrospective Analysis for Research and Applications (MERRA), and in its second version (MERRA-2) and found an improved representation of the QBO in MERRA-2, mainly due to the retuning of the gravity wave drag (GWD) parameterization in the model, amplifying it in the tropics by nearly a factor of 8 compared to MERRA (Molod et al. 2015).

The ECMWF ERA-Interim (ERA-I) reanalysis has also been used extensively for assessing various aspects of the QBO, including its momentum forcing by the equatorial waves (e.g., Ern et al. 2014; Kim and Chun 2015; Schenzinger et al. 2017). ERA-I has been found to be quite reliable in the tropical lower and middle stratosphere and to have a particularly good representation of planetary waves (e.g., Ern and Preusse 2009a,b).

Recently, Anstey et al. (2020) examined the representation of the QBO in various reanalyses and found that there is a good agreement between the modern products on the evolution of the QBO-related zonal wind variations and the relative contributions of the various tropical waves to the forcing of the QBO. Here we revisit the QBO using ERA5, the fifth generation of ECMWF atmospheric reanalyses (Hersbach et al. 2020). This paper (Part I) provides a general overview of the structure and dynamics of the QBO, as represented in ERA5 and an indication of the ways in which the representation is improved in ERA5 relative to ERA-I. The goal is to gain a deeper understanding of the QBO dynamics and momentum budget.

ERA5 improves upon ERA-I in several respects. Its considerably higher spatial resolution in both the horizontal and vertical makes it possible to resolve a broader spectrum of atmospheric waves in ECMWF\textquotesingle s Integrated Forecasting System (IFS). With its higher vertical resolution it is also possible to represent the process of wave-mean flow interaction more realistically (Dunkerton 1997). Numerous previous modeling studies have shown that increasing the resolution increases the Eliassen-Palm flux (EP flux) divergence (i.e., the momentum forcing by the resolved waves) and that increasing the vertical resolution has a much larger effect than increasing the horizontal resolution (Holt et al. 2016, 2020).

More detailed descriptions of ERA5 and the methods used in this study are presented in section 2. In section 3, we document the structure of the QBO-related axisymmetric fields, and discuss the interactions of the QBO with semiannual oscillation (SAO), and the Brewer-Dobson circulation (BDC). We also describe the seasonality of the QBO in this section. Section 4 presents an overview of the momentum budget of the QBO. It is shown that the advection of westerly momentum acts to damp the QBO. Hence, the wave forcing is responsible, not only for the downward propagation, but also for maintaining its amplitude. A more detailed documentation of the contributions of different modes of atmospheric waves to driving the QBO is presented in Part II. Section 5 of Part I compares the performance of ERA5 and ERA-I in representing the QBO. While the mean fields are very similar in the two reanalyses, the resolved wave forcing is shown to be substantially stronger in ERA5 than in ERA-I. Section 6 presents a few concluding remarks.

\section{Data and methodology}

The ERA5 global reanalysis for 1979 to present is the primary basis for this study. It is produced using the IFS cycle 41r2 with four-dimensional variational (4-DVar) data assimilation. It has a horizontal resolution of $\sim$31 km ($\sim$0.28\textdegree ) and 137 hybrid sigma-pressure (model) levels extending up to 0.01 hPa (80 km). The improved spatial and temporal resolution of the ERA5 reanalysis relative to ERA-I allows for a better representation of convective updrafts, gravity waves, tropical cyclones, and other meso- to synoptic-scale features (Hoffmann et al. 2019; Hersbach et al. 2020), Furthermore, the use of an updated and significantly improved numerical weather prediction model and the data assimilation system compared to earlier reanalyses, along with the incorporation of data from additional sources, have improved the predictive skill of the forecast model used in producing ERA5, making it possible to use a smaller data assimilation increment (Hoffmann et al. 2019).

In this study, we use the high resolution (HRES) version of ERA5 and retrieve the 6-hourly  data at 0.5\textdegree  $\times$ 0.5\textdegree horizontal sampling on model levels, compared to 1.5\textdegree  in ERA-I. Since the effective resolution of weather prediction models is typically on the order of 4 $\Delta x$, where $\Delta x$ is the native resolution of the model (e.g., Skamarock et al. 2014), ERA5 should be capable of resolving equatorial waves with zonal wavelengths as short as 1\textdegree  longitude, equivalent to zonal wavenumber 360. Between 100 and 1 hPa, the ERA5 has 41 vertical levels with a resolution of $\sim$300 m at 100 hPa, decreasing gradually to $\sim$1.5 km at 1 hPa, as compared to 20 vertical levels in ERA-I with a resolution of $\sim$1.1 km at 100 hPa decreasing to $\sim$2.5 km at 1 hPa.

For evaluating the terms in the zonal momentum budget, we use the zonal mean zonal momentum equation in the transformed Eulerian mean (TEM) form with spherical geometry and log-pressure coordinates (Andrews et al. 1987):
\smallskip

$\bar u_t = \bar v^*[f-(a \cos\phi)^{-1}(\bar u \cos\phi)_\phi] - \bar w^*\bar u_z \\
\hspace*{25pt} + (\rho_0 a \cos\phi)^{-1}\nabla . \mathbf{F} + \bar X \hfill (1) $
\smallskip

Here, $\bar u_t$ is the tendency of the zonal mean zonal wind, $f$ the Coriolis frequency, $a$ is the radius of Earth, $\phi$ is latitude, and $\rho_0  (z)$ is the reference density. $\bar v^*$ and $\bar w^*$ are the meridional and vertical components of the residual circulation, respectively, defined as
\smallskip

$\bar v^* = \bar v - \rho_0^{-1} (\rho_o \overline{v'\theta'} / \bar\theta_z)_z \hfill (2) $

$\bar w^* = \bar w + (a \cos\phi)^{-1} (\cos \phi \overline{v'\theta'} / \bar\theta_z)_\phi \hfill (3) $

\smallskip

$ \mathbf{F}$ is the EP flux, whose meridional and vertical components are
\smallskip

$F^{(\phi)} = \rho_0 a \cos \phi (\bar u_z \overline{v'\theta'} / \bar\theta_z - \overline{v'u'}) \hfill (4)$

$F^{(z)}  = \rho_0 a \cos \phi \{[f - (a \cos \phi)^{-1} (\bar u \cos \phi)_\phi]  \overline{v'\theta'} / \bar\theta_z \\
\hspace*{35pt} - \overline{w'u'}\} \hfill (5) $

\begin{figure*}[t]
\centerline{\includegraphics[width=\textwidth]{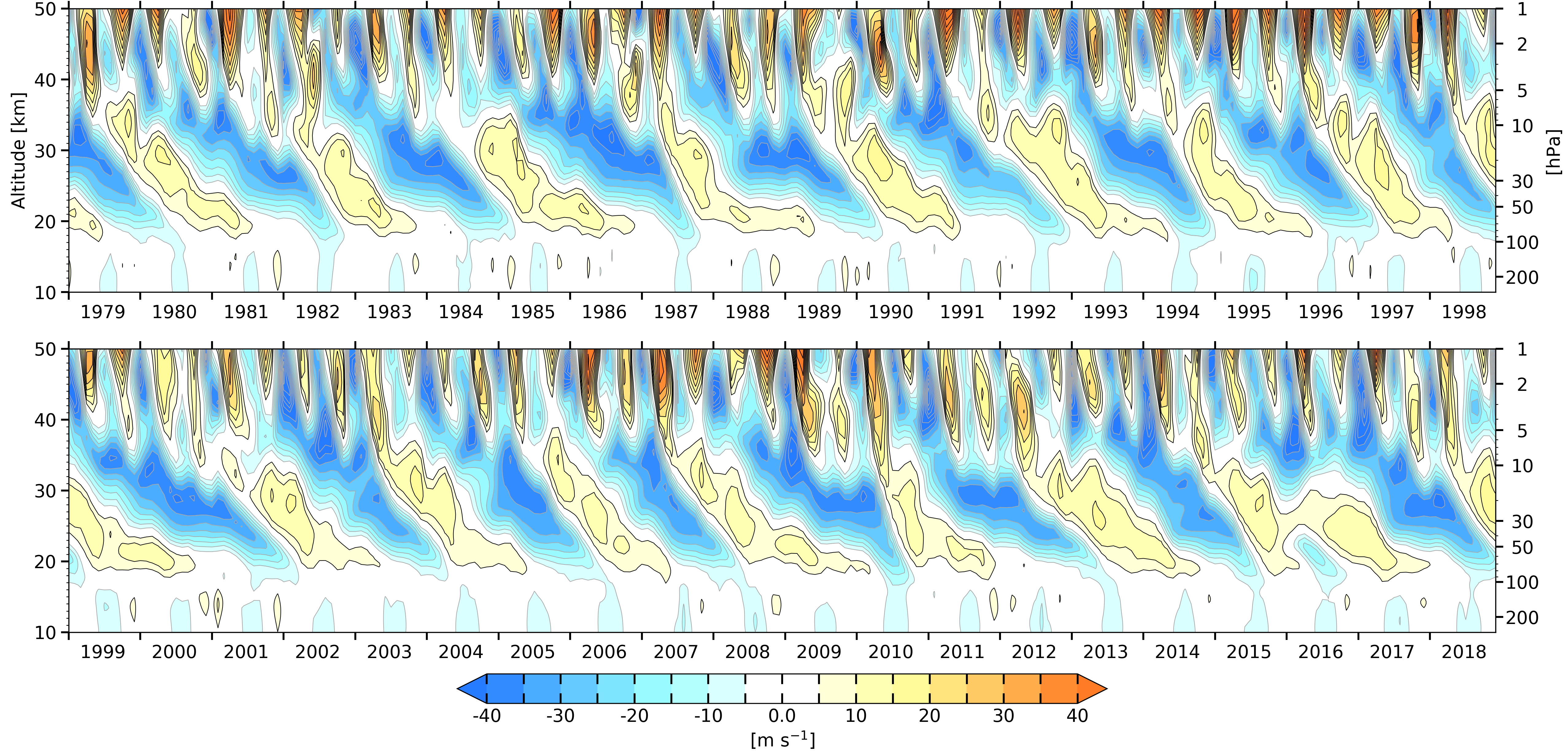}}

\caption{Time-height section of monthly zonally averaged zonal wind averaged over 5\textdegree  N/S. The tick marks on the x-axis represent January 15 for each year.} \label{fig1}
\end{figure*}

In these expressions $\theta$ is potential temperature and the subscripts $(\phi)$ and $(z)$ indicate differentiation in the meridional and vertical directions, respectively. Overbars denote zonal means, and primes denote deviations from zonal means. Above 73 hPa ($\sim$18 km), model levels correspond to pressure levels. Hence, in effect, pressure is used as the vertical coordinate in evaluating the terms in Eq. (1).

The divergence of the EP flux represents the momentum forcing by the waves that are explicitly resolved in the reanalysis. The residual zonal wind tendency $\bar X$, represents the zonal forcing associated with subgrid-scale waves in the reanalysis, including mesoscale gravity waves forced by dynamical processes such as convection, frontogenesis, processes involving jet stream fine structure, and flow instabilities. These waves have to be parametrized using schemes that are subject to large uncertainties. One of the objectives of this paper is to assess how well they are represented.

\section{Structure of the axisymmetric fields}

The time-height section of the monthly mean zonally averaged zonal wind, averaged over 5\textdegree  N/S, shown in Fig. 1, reveals many of the essential dynamical properties of the QBO. A succession of westerly and easterly wind regimes descends through the equatorial stratosphere from $\sim$35 km ($\sim$6 hPa) down to the vicinity of the equatorial cold point tropopause at an average rate of $\sim$1 km per month in agreement with previous studies (e.g., Baldwin et al. 2001 and references therein). The descent rate is noticeably faster and more continuous for the westerly shear zones (i.e., layers with positive $\partial u / \partial z $). As a result, the westerly regimes last noticeably longer than the easterly regimes at 50 hPa and below, while the reverse is true at 10 hPa and above. The peak winds in the easterly regimes are more than twice as strong as those in the westerly regimes. This disparity is consistent with the fact that the stratospheric mean climatology is characterized by deep layers of off-equatorial easterlies with peak values $\sim$14 m s$^{-1}$  centered $\sim$12\textdegree  N/S. Apart from the short lapse in 2016 (see Part II for more details), the zonal wind variations are quasi-periodic, with the length of full cycles ranging from about 22 to 34 months, with an average of $\sim$28 months, consistent with the label "quasi-biennial oscillation", a term coined by Angell and Korshover (1964).

In the season from July to February the QBO often stalls for several months when it is in the phase of its cycle with the peak easterly winds in the 30-50 hPa layer and westerlies in the vicinity of the cold point tropopause $\sim$100 hPa. These irregularities in the cycle have been attributed to seasonal variations in the wave forcing. During the boreal summer the tropical upper tropospheric zonal winds are more easterly and less favorable for the upward dispersion of westward propagating waves (Maruyama 1991; Krismer et al. 2013), and during the boreal winter the stronger equatorial upwelling tends to slow the downward phase propagation of the QBO (Kinnersley and Pawson 1996; Hampson and Haynes 2004; Dunkerton 2016).

It is evident from Fig. 1 that the QBO vanishes as it approaches the tropical tropopause layer, even though the vertically propagating waves that drive the QBO originate in the troposphere several kilometers below. Using MERRA-2 and ERA-I reanalysis, Match and Fueglistaler (2019) showed that mean-flow damping due to horizontal momentum flux divergence weakens the QBO amplitude in this region.

\begin{figure*}[t]
\centerline{\includegraphics[width=25pc]{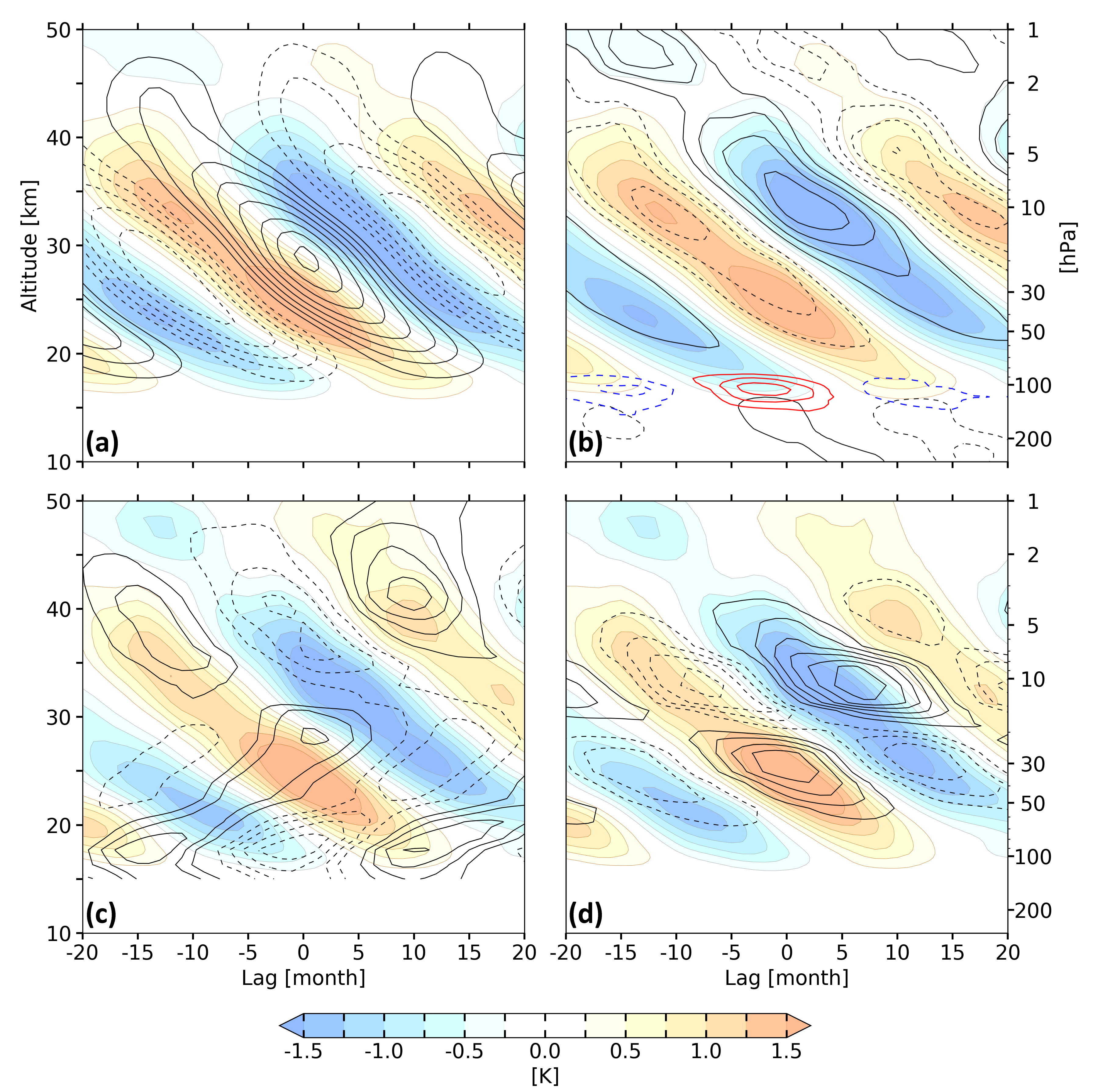}}

\caption{Time (lag)-height section created by regressing time series of zonally averaged variables (contours) and temperature (colored shading), averaged over 5\textdegree  N/S, at different altitudes and time lags upon zonally averaged zonal wind at the 29 km ($\sim$14 hPa) level using monthly data. The contours are (a) zonal wind, contour interval 2.5 m s$^{-1}$, (b) vertical velocity (black), contour interval 0.075 mm s$^{-1}$, and cloud fraction (red and blue), contour interval 0.75\%, (c) water vapor, contour interval 0.04 ppmv (not shown below 15 km), and (d) ozone, contour interval 0.075 ppmv. The zero contours are omitted, and negative contours are dashed. The top panels are based on a 40-year record (1979 to 2018), while the bottom panels are based on a 14-year record (2004 to 2018).} \label{fig2}
\end{figure*}

The flow in the upper stratosphere and mesosphere exhibits a pronounced semiannual oscillation, characterized by westward wind maxima a few weeks after the solstices and eastward wind maxima a few weeks after the equinoxes, as first documented by Reed (1962, 1966). It has been suggested that the westerly phase of the SAO is due to eastward propagating waves, in particular shorter-period Kelvin waves with high phase speed, while the easterly phase is driven by a combination of planetary wave forcing and the meridional advection of angular momentum in the upper branch of the Brewer-Dobson circulation (BDC), consisting of rising motions in the summer hemisphere, a cross-equatorial drift, and sinking in the winter hemisphere (Holton and Wehrbein 1980).

It is evident from Fig. 1 that the QBO and SAO share the same westerly shear zones at altitudes between about 35-40 km (Smith et al. 2017). Once in every 4 to 6 semiannual cycles, the westerly wind shear zone at the base of the SAO ($\sim$40 km) begins to propagate downward, initiating the descent of a new westerly shear zone of the QBO. Hence, the arrival of successive westerly QBO regimes at 30 km tends to occur at multiples of 6 months, as hypothesized by Lindzen and Holton (1968) and verified in observations by Garcia et al. (1997). Theoretical and idealized model studies and laboratory experiments suggest that QBO westerly shear zones can originate in the upper stratosphere in the absence of preexisting SAO westerly shear zones (Holton and Lindzen 1972; Plumb 1977; Plumb and McEwan 1978; Mayr et al. 2010). Yet, within the layer in which the SAO is active, the deposition of eastward momentum by breaking waves tends to be focused in its westerly shear zones, enabling them to continue propagating downward far beyond the layer in which the SAO is dominant. In effect, what was an SAO westerly shear zone becomes a QBO westerly shear zone. These transitions occur in the phase of the QBO when westerly wind regimes are present in the lower stratosphere (Wallace 1973; Dunkerton and Delisi 1997; Garcia et al. 1997). As the westerlies weaken, the lower stratosphere becomes more permeable to the waves that are producing an upward transport of eastward momentum, because fewer of them encounter their critical level (i.e., the level at which their wave phase speed is close to the zonal wind speed) and break. With more eastward momentum reaching the upper stratosphere, the SAO westerly shear zones are forced more strongly, enabling them to penetrate deeper into the lower stratosphere (e.g., Krismer et al. 2013).

Figure 2 summarizes the characteristics of the QBO as they appear in equatorial time-height sections for various variables. It shows time (lag) vs. height sections created by regressing monthly time series of zonally-averaged zonal wind and other variables at different levels and time lags upon a common reference time series of monthly equatorial zonally-averaged zonal wind at the 29 km ($\sim$14 hPa) level, scaled by its one standard deviation. The results are not sensitive to the choice of reference level. The annual cycle is removed by subtracting the 40-yr average of each month. For reference, the patterns in all panels are superimposed upon the corresponding QBO-related temperature anomalies as inferred from linear regression. Consistent with Fig. 1, the QBO signal in zonal wind propagates downward with a period of $\sim$28 months. Consistent with the scale analysis of Reed (1962), the zonal wind and temperature fields are seen to be in thermal wind balance, with positive temperature anomalies in the westerly shear zones and vice versa. The QBO-related temperature anomalies are strongest near 30-50 hPa, with values ranging up to $\pm$1.5 K (Reid 1994), and they weaken as they propagate downward, to on the order of $\pm$0.5 K near the cold point tropopause ($\sim$100 hPa) (Randel et al. 2000).

In the absence of dynamically induced heating/cooling, radiative relaxation would damp these temperature anomalies on a time scale of weeks (Hartmann et al. 2001; Randel et al. 2002). They are maintained, as required for thermal wind balance, by the adiabatic warming and cooling associated with the axially symmetric vertical motions, downward in the relatively warm westerly shear zones and upward in the cold easterly shear zones, as indicated by the black contoured field in Fig. 2b. For reference, the climatological mean ascent rate in the BDC is $\sim$0.15 mm s$^{-1}$ at the 60 hPa level and $\sim$0.25 mm s$^{-1}$ at the 30 hPa level (Lin et al. 2013). The descent in the westerly shear zones within 5 degrees of the equator is thus strong enough to largely compensate and even temporarily interrupt the ascent associated with the BDC. An anomalous upward vertical motion in the troposphere is also evident in Fig. 2b during the easterly phase of the QBO, implying deeper and/or more convection. Several studies have suggested that the QBO is able to modulate the tropical deep convection by perturbing the state of the lower stratosphere and upper troposphere (see Nie and Sobel 2015 and reference therein).

Vestiges of the QBO-related temperature perturbations penetrate downward deep enough into the tropical tropopause layer (TTL) to modulate the areal coverage of high, thin cirrus clouds (Tseng and Fu 2017), here defined as clouds with bases higher than 14.5 km. Negative temperature anomalies favor higher relative humidity and consequently more extensive cloud cover and vice versa, as indicated by red and blue contours in Fig. 2b, based on 12 years (2006 to 2018) of observations from the Cloud-Aerosol Lidar and Infrared Pathfinder Satellite Observations (CALIPSO).

The concentrations of many trace constituents are also affected by the QBO. Schoeberl et al. (2008) documented the QBO signals in multiple trace gases based on HALOE and Aura MLS observations. Here, we consider only water vapor and ozone based on a 14-year record (August 2004 to December 2018) of MLS observations. 

\begin{figure*}[t]
\centerline{\includegraphics[width=32pc]{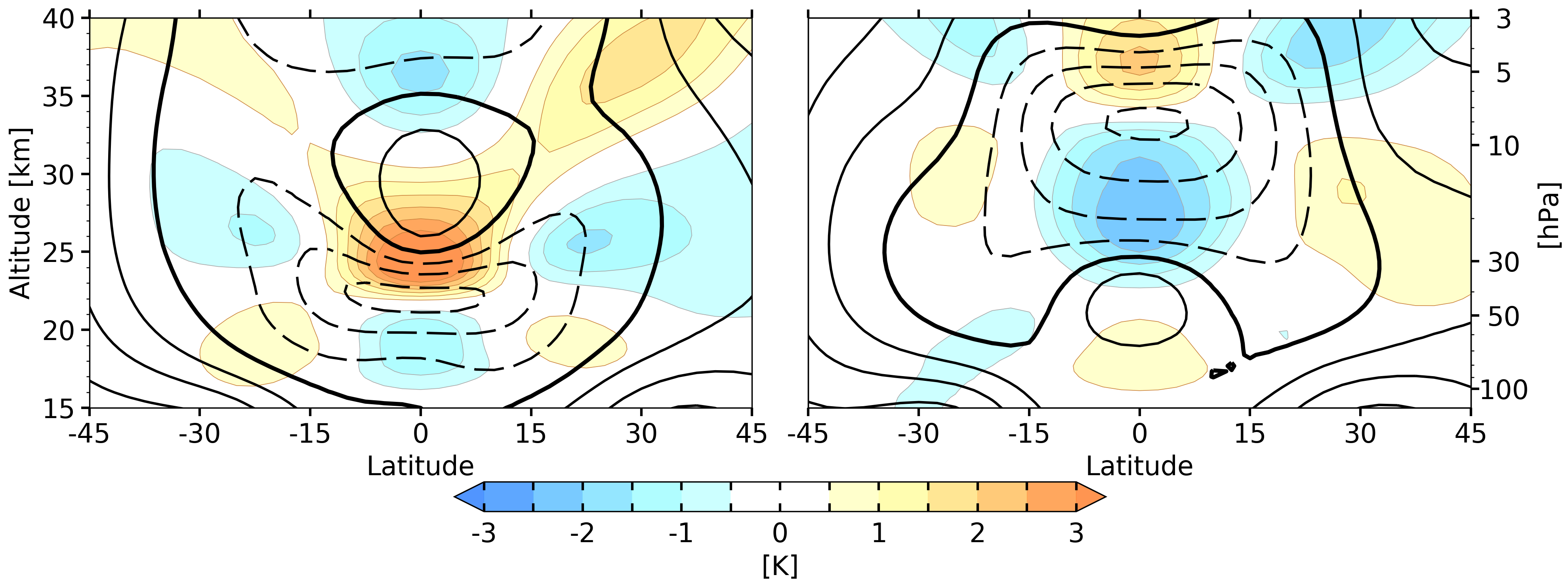}}

\caption{Composites of temperature anomalies (colored shading) and zonal wind (contours) in descending westerly (left) and easterly (right) shear zones of the QBO, as defined in the text. Note that the zonal wind represents the total field, not just the anomalies. Contour interval 7.5 m s$^{-1}$, westerlies solid, easterlies dashed, zero contour bolded.} \label{fig3}
\end{figure*}

It is well recognized that the temperature of the tropical cold point controls the amount of water vapor entering the lower stratosphere (Brewer 1949). The ascent of alternating moist and dry layers formed during different phases of the annual cycle in cold point temperature has come to be referred to as "atmospheric tape recorder" (Mote et al. 1996). It is evident from Fig. 2c that QBO-related temperature variations also modulate the tropical tropopause temperatures and hence water vapor concentrations of air entering the stratosphere in a similar manner, with more water vapor entering the stratosphere when QBO-related temperature anomalies are warm near the cold point at around 100 hPa. That the QBO-related anomalies in water vapor mixing ratio maintain their integrity as they ascend through the lower stratosphere from the cold point up to $\sim$30 km at approximately the rate of the background mean upwelling in the BDC, is consistent with the tape recorder analogy, as articulated, for example, by Fueglistaler and Haynes (2005). In the upper stratosphere, the QBO water vapor anomalies are approximate mirror images (with opposite sign) of those in methane (not shown), because the oxidation of methane is the main source of the water vapor above 35 km (Randel et al. 1998). Considering that the rate of methane oxidation is temperature-dependent, QBO-related water vapor variations in the upper stratosphere are likely due to QBO-induced perturbations in both temperature and advection. That the westerly and easterly wind regimes of the QBO propagate downward through these ascending moist and dry layers proves that they must be propelled by downward propagating sources of momentum, as discussed in the next section.

The QBO signal in the equatorial ozone, shown in Fig. 2d, reveals the existence of separate lower stratospheric (20-27 km) and middle stratospheric (30-37 km) regimes, with a transition at around 28 km (e.g., Schoeberl et al. 2008 and references therein). Below 28 km, the chemical lifetime of ozone is relatively long compared with dynamical transport processes and hence it may be regarded as a long-lived, passive tracer. Because the background ozone mixing ratio increases rapidly with height in this layer, the QBO-related ozone anomalies are positive in regions of descent and vice versa. Above 28 km ozone chemical lifetime is much shorter, limiting the direct impact of advection. At these altitudes, the key catalytic cycles controlling ozone loss are linked to reactive nitrogen in the form of nitrogen oxides (NOy) (Park et al. 2017). Hence, the QBO-induced perturbations in chemical loss rate due to variations in temperature and NOy transport are the main cause of the QBO-related ozone variations above 28 km, where its concentration is largely controlled by the ambient temperature (Fleming et al. 2002).

Meridional cross sections of the total zonal wind and temperature anomalies are shown in Fig. 3 for the QBO westerly (left) and easterly (right) shear zone composites, constructed by averaging variables for the months when the zero-wind line of zonally averaged zonal wind over the equator in a descending westerly or easterly shear zones reaches a prescribed reference level, here taken to be 25 hPa ($\sim$25 km). The months used in constructing each composite are indicated in Fig. S1. Since linear regression is not employed in creating the composites, the anomaly fields are not constrained to be equal and opposite in contrasting composites. In contrast to the zonal wind anomalies shown in many previous studies (e.g., Randel et al. 1999), the easterly wind regimes in the total wind field are much wider than the westerly regimes, consistent with the mean climatology, as noted above. The disparity could simply be a reflection of the background climatology, as suggested by Randel and Wu (1996). But if the structure and the peak wind speeds in westerly and the easterly wind regimes are shaped by the spectrum of waves dispersing upward from below, the mean climatology might be shaped by the QBO. Another distinctive feature of the cross sections shown in Fig. 3 is that the westerly regimes exhibit U-shaped bases, while the easterly regimes exhibit flatter bases, as pointed out by Hamilton (1984). This difference also appears to be a consequence of wave-mean flow interaction, as discussed in Section 4.b.

\begin{figure}[t]
\centerline{\includegraphics[width=19pc]{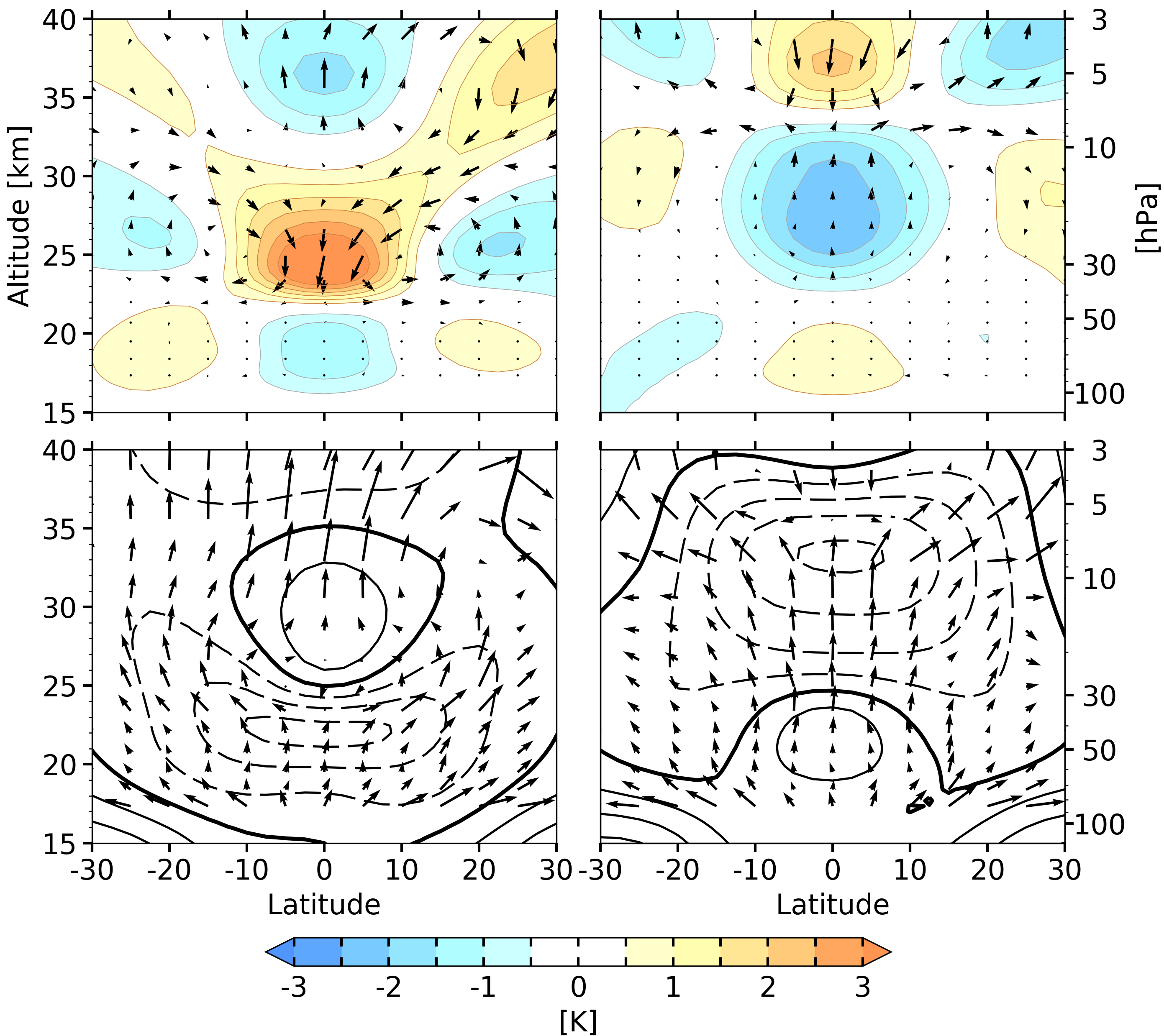}}

\caption{Composites of temperature anomalies (colored shading), zonal wind (contours), and TEM mean meridional circulation (MMC) (arrows) for descending westerly (left) and easterly (right) shear zones of the QBO. The QBO-induced MMC is shown in the top panels and the total MMC, which includes the climatological mean, is shown in the bottom panels. Contour interval 7.5 m s$^{-1}$, westerlies solid, easterlies dashed, zero contour bolded. The longest horizontal components of the arrows ($\bar v^*$) correspond to 0.3 m s$^{-1}$, while the longest vertical components ($\bar w^*$) correspond to 1 mm s$^{-1}$.} \label{fig4}
\end{figure}

Figure 4 documents the QBO-induced and total mean meridional circulations (MMC). The QBO-induced MMC composite patterns shown in the top panels, with descent in the westerly shear zones and ascent in the easterly shear zones are in qualitative agreement with the circulations postulated by Reed (1964) and Wallace (1967) on the basis of simple dynamical and thermodynamic arguments. The vertical advection associated with QBO-induced MMC thus speeds the descent of westerly wind regimes while it impedes the descent of easterly regimes. The temperature field in the QBO is maintained against radiative relaxation by the QBO-induced MMC. In this sense, the QBO-induced MMC can be viewed as a consequence of radiative relaxation, which drives the wind and temperature fields out of thermal wind balance. In effect, the QBO-induced MMC creates and sustains the QBO-related temperature perturbations while it damps the QBO-related zonal wind perturbations, as discussed in section 4.a.

The total TEM MMC fields constructed by adding the perturbations in the top panels of Fig. 4 to the climatological annual mean BDC are shown in the bottom panels. Within 5\textdegree  of the equator the descent in the westerly shear zones is strong enough to cancel the BDC-related upwelling. The flow in the BDC is deflected around their flanks. In the seasonally dependent composites, the flow is deflected preferentially into the winter hemisphere (not shown). In the easterly shear zones the upwelling is enhanced. Hence, the collapse of westerly wind regimes as they approach the tropopause is attended by a breakthrough of the BDC-related upwelling, increasing the concentrations of tropospheric trace constituents, and decreasing the concentrations of ozone and other stratospheric trace constituents.

\begin{figure}[t]
\centerline{\includegraphics[width=17pc]{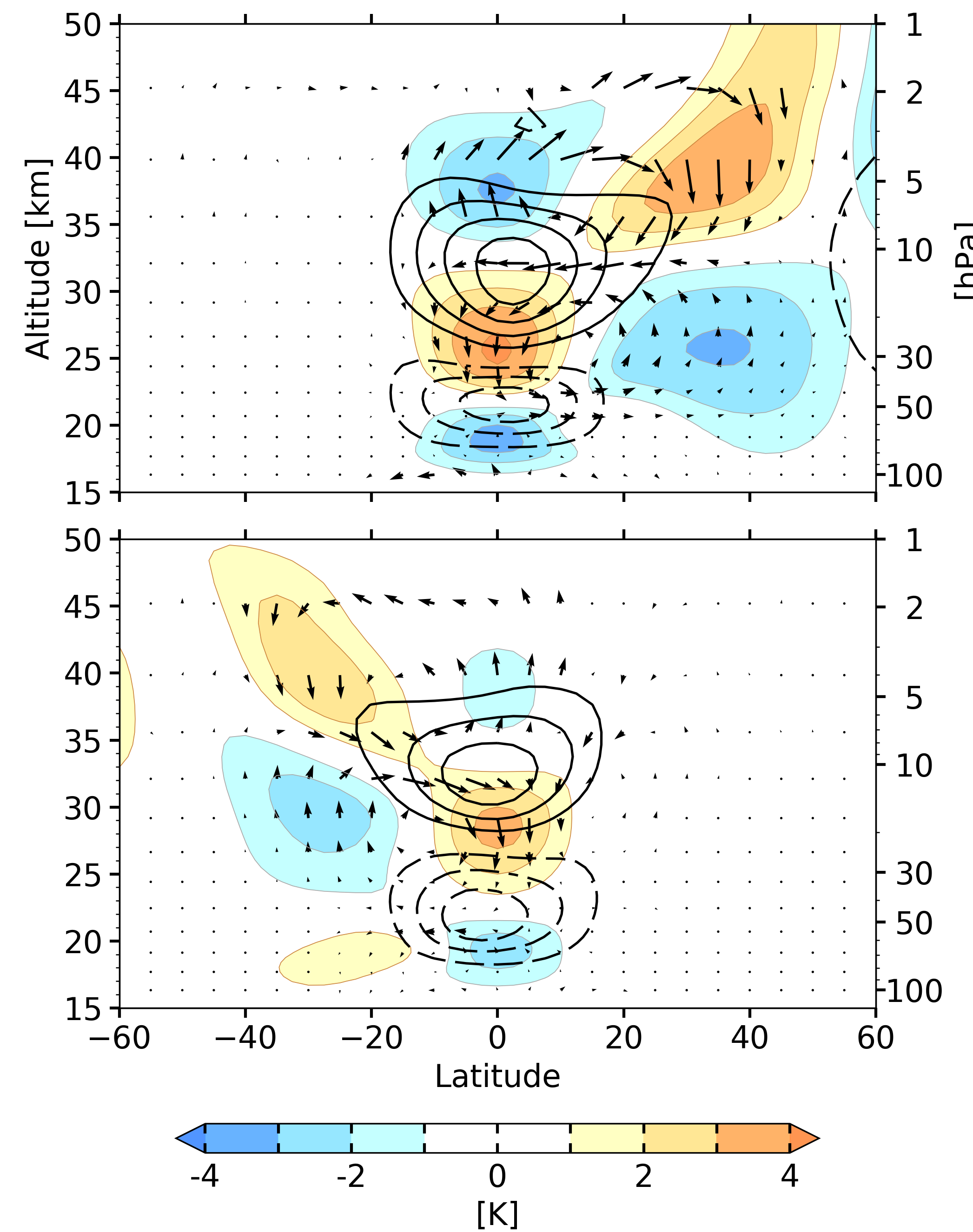}}

\caption{Meridional cross sections of temperature (colored shading), MMC (arrows) and zonal wind (contours) for easterly minus westerly composites as defined by the magnitude of zonally averaged zonal wind > 5 m s$^{-1}$ at the 50 hPa level for DJF (top), and JJA (bottom). Contour interval 7.5 m s$^{-1}$, westerlies solid, easterlies dashed, zero contour omitted. The longest horizontal components of the arrows ($\bar v^*$) correspond to 0.4 m s$^{-1}$, while the longest vertical components ($\bar w^*$) to 0.6 mm s$^{-1}$.} \label{fig5}
\end{figure}

\begin{figure*}[t]
\centerline{\includegraphics[width=\textwidth]{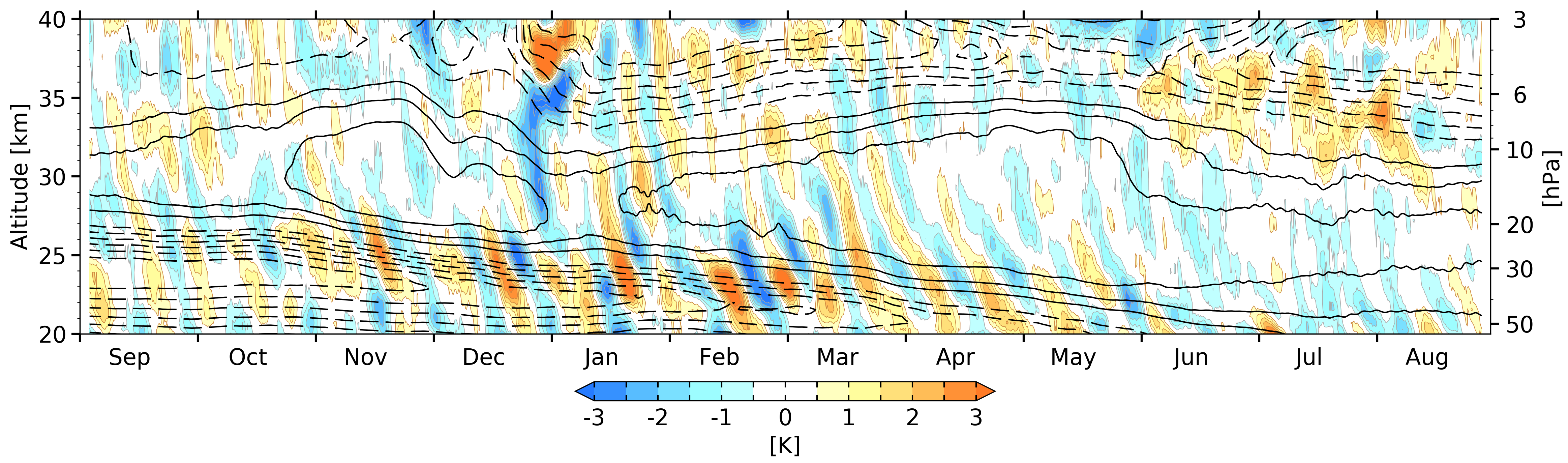}}

\caption{Time-height section of the wave component of the temperature field (colored shading) for the equatorial grid point at the longitude of the Date Line (0\textdegree , 180\textdegree ) in calendar year 2012-2013. The contours indicate equatorial zonally averaged zonal wind, contour interval 5 m s$^{-1}$, westerlies solid, easterlies dashed, zero contour omitted. The temperature perturbations are smoothed by a 5-day running mean.} \label{fig6}
\end{figure*}

From Figs. 3 and 4 it is readily apparent that the QBO-related temperature anomalies and the corresponding MMCs are nearly equatorially symmetric when estimated on the basis of year-round data. However, when the westerly and easterly shear zone composites are constructed based on seasonal mean data, important equatorial asymmetries emerge. The composites shown in Fig. 5 are constructed separately for 3-month seasons December-February (DJF) and June-August (JJA). The criterion for selecting the months included in the easterly composite is that zonally-averaged zonal wind at the 50 hPa level be easterly and in excess of -5 m s$^{-1}$ and the same criterion with reversed sign is used for the westerly composite. Since the composites for westerly and easterly phases are similar but with the sign reversed, only their difference (i.e., easterly minus westerly) is shown in Fig. 5. The subtropical/midlatitude centers of action in the QBO-related temperature and MMC fields are confined to the winter hemispheres. Concentrations of tracers transported by QBO-induced MMC, such as ozone, also exhibit an analogous seasonality, with much larger poleward transports in the winter hemisphere (Randel and Wu 1996). The seasonality of the QBO-induced MMC can be related to the seasonally varying hemispheric asymmetries in the BDC, whose poleward branch and associated planetary Rossby wave driving is concentrated in the winter hemisphere (Holton 1989). A likely explanation is that the QBO-induced mean meridional motions can extend into middle latitudes only in the presence of Rossby wave-breaking and as noted by Hitchman and Huesmann (2007), at stratospheric levels Rossby waves are present only in the winter hemisphere.

\section{The momentum budget }

The smoothness of the zonal wind field in Fig. 1 is due to the zonal averaging, which effectively removes the fluctuations associated with the passage of atmospheric waves. The colored shading in Fig. 6 shows a time-height section of the resolved wave component (i.e. departures from the zonal mean) of temperature at an equatorial grid point on the Date Line during a year-long interval in which a westerly shear zone, indicated by the zonal wind contours, is descending through the lower stratosphere. To simplify the signature of the waves, the temperature perturbations are smoothed by applying a 5-day running mean. They are wavelike with periods on the order of 10-15 days and downward propagating, and particularly well organized within the slowly descending westerly shear zone.

Equatorially-trapped waves with a variety of vertical and horizontal wavelengths and phase speeds, excited mainly by convection in the tropical troposphere, disperse upward into the stratosphere, producing transports of momentum and heat. As the waves approach their critical levels, they either break or are radiatively damped (Holton and Lindzen 1972) and deposit their zonal momentum to the mean flow (Booker and Bretherton 1967). The absorption of the waves induces a downward displacement of the shear zone and this, in turn, causes the waves to break and deposit their momentum at progressively lower levels, so that the shear zone continues to descend. In Fig. 6 the westerly shear zone descends from $\sim$27 km to below $\sim$20 km over the course of one year. As shown in Part II, much of the regular 10-15 day signal descending with the shear zone can be associated with Kelvin waves.

The resolved wave forcing is only part of the QBO momentum budget. Advection by the mean meridional circulations as well as unresolved wave forcing also contribute to the total zonal wind tendency. In what follows, we will use the TEM equations summarized in Section 2, in an effort to gain a better understanding of the role of various terms in Eq. (1) in the maintenance and downward propagation of the QBO.

\begin{figure*}[t]
\centerline{\includegraphics[width=27pc]{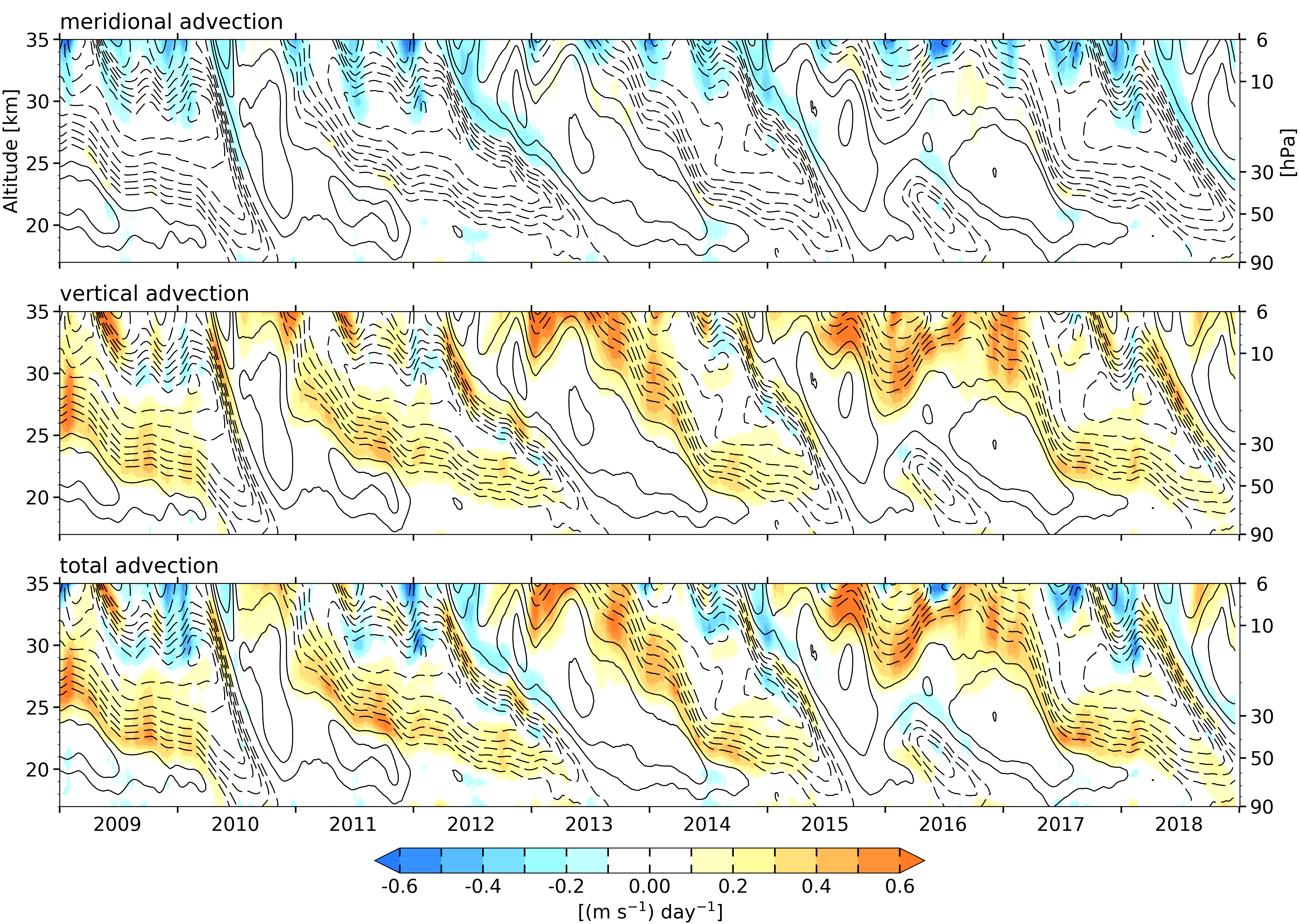}}

\caption{Time-height sections of zonal wind tendency (colored shading) due to the advection terms in Eq. (1) as indicated, superimposed upon the zonally averaged zonal wind (contours), averaged over 5\textdegree  N/S. Contour interval 5 m s$^{-1}$, westerlies solid, easterlies dashed, zero contour omitted. Based on 6-hourly data, smoothed by a 30-day running mean.} \label{fig7}
\end{figure*}

\begin{figure}[t]
\centerline{\includegraphics[width=16pc]{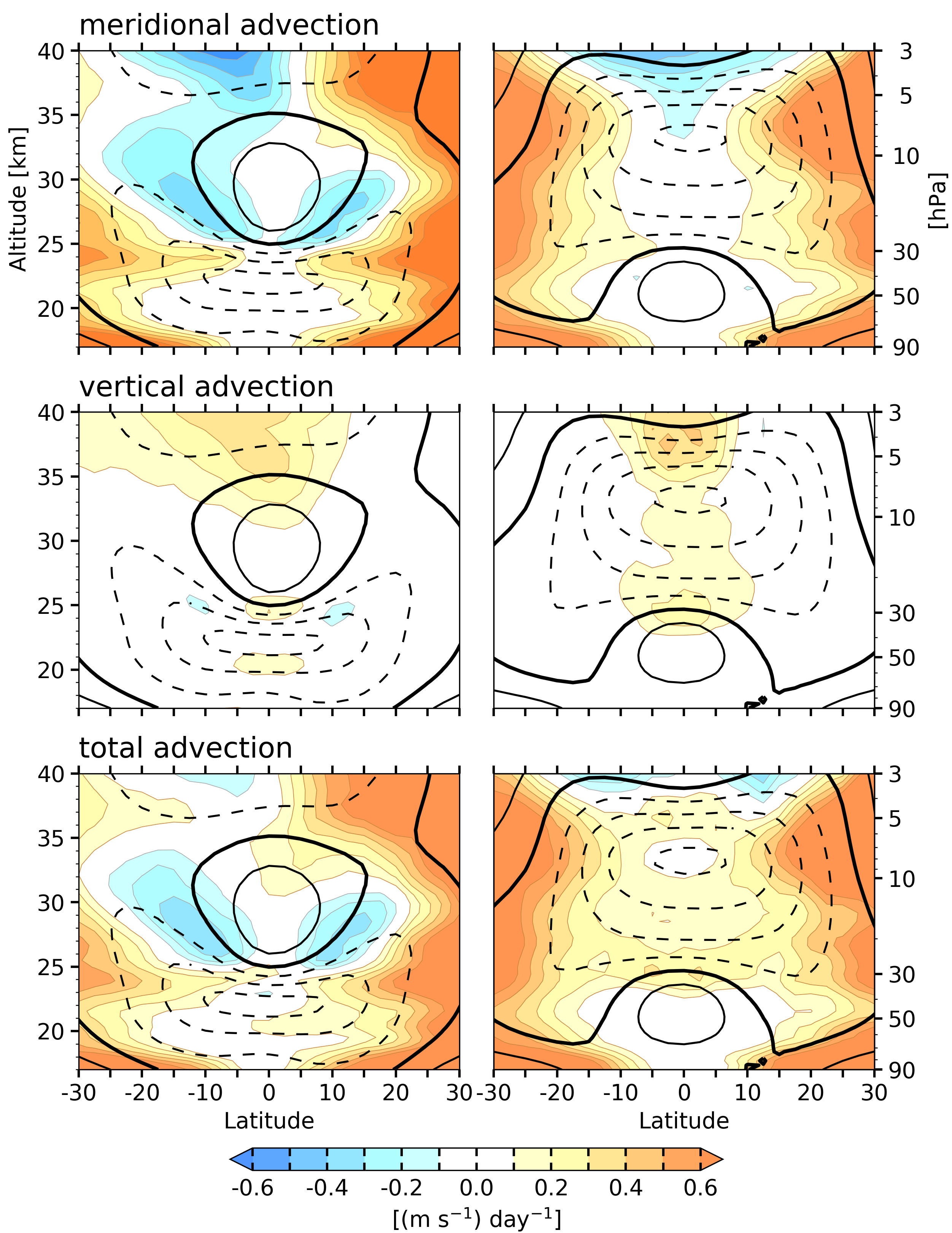}}

\caption{Composites for descending westerly (left) and easterly (right) shear zones of the QBO. The colored shading shows the zonal wind tendency due to the advection terms in Eq. (1) as indicated, superimposed upon the zonally averaged zonal wind (contours). Contour interval 7.5 m s$^{-1}$, westerlies solid, easterlies dashed, zero contour bolded.} \label{fig8}
\end{figure}

\subsection{The role of advection}

Figure 7 shows time-height sections of the advection terms in Eq. (1) averaged over 5\textdegree  N/S for a 10-yr interval and Fig. 8 shows their distribution in the meridional plane in composites for westerly and easterly shear zones, constructed in the same manner as Fig. 3. In both figures the zonal wind tendencies are superimposed upon the distribution of zonal wind itself. The meridional advection, the first term on the right-hand side of Eq. (1), is small in the equatorial time-height section. The most notable feature is the westward acceleration that occurs regularly at 6-month intervals above $\sim$30 km. It peaks at the times of the solstices, when air is flowing across the equator from the summer hemisphere to the winter hemisphere in the upper branch of the BDC (Delisi and Dunkerton 1988). Holton and Wehrbein (1980) proposed that the seasonally reversing cross-equatorial flow at these levels plays a prominent role in forcing the SAO. The meridional advection term in Fig. 7 also produces weak westward acceleration within the westerly shear zones. The strongest influence of meridional advection term is off the equator, where the $f \bar v^*$ term makes the dominant contribution, as shown in the top panel of Fig. 8. It tends to damp the QBO by weakening the easterly wind regimes along their poleward flanks and by eroding the westerly wind regimes along their bottom edge.

The vertical advection term, the second term on the right-hand side of Eq. (1), shown in the middle panels of Figs. 7 and 8 also tends to weaken the easterly regimes along the equator, but off the equator it is overshadowed by the meridional advection term. The net effect of the two terms, shown in the bottom panels, is to damp the QBO (e.g., as noted by Dunkerton 1997) without contributing appreciably to its downward propagation. That they should have a net damping effect is consistent with the energetics of the QBO and BDC, whose mean meridional motions are known to be thermally indirect, converting kinetic energy to available potential energy, which is subject to radiative damping (Wallace 1967).

\begin{figure*}[t]
\centerline{\includegraphics[width=28pc]{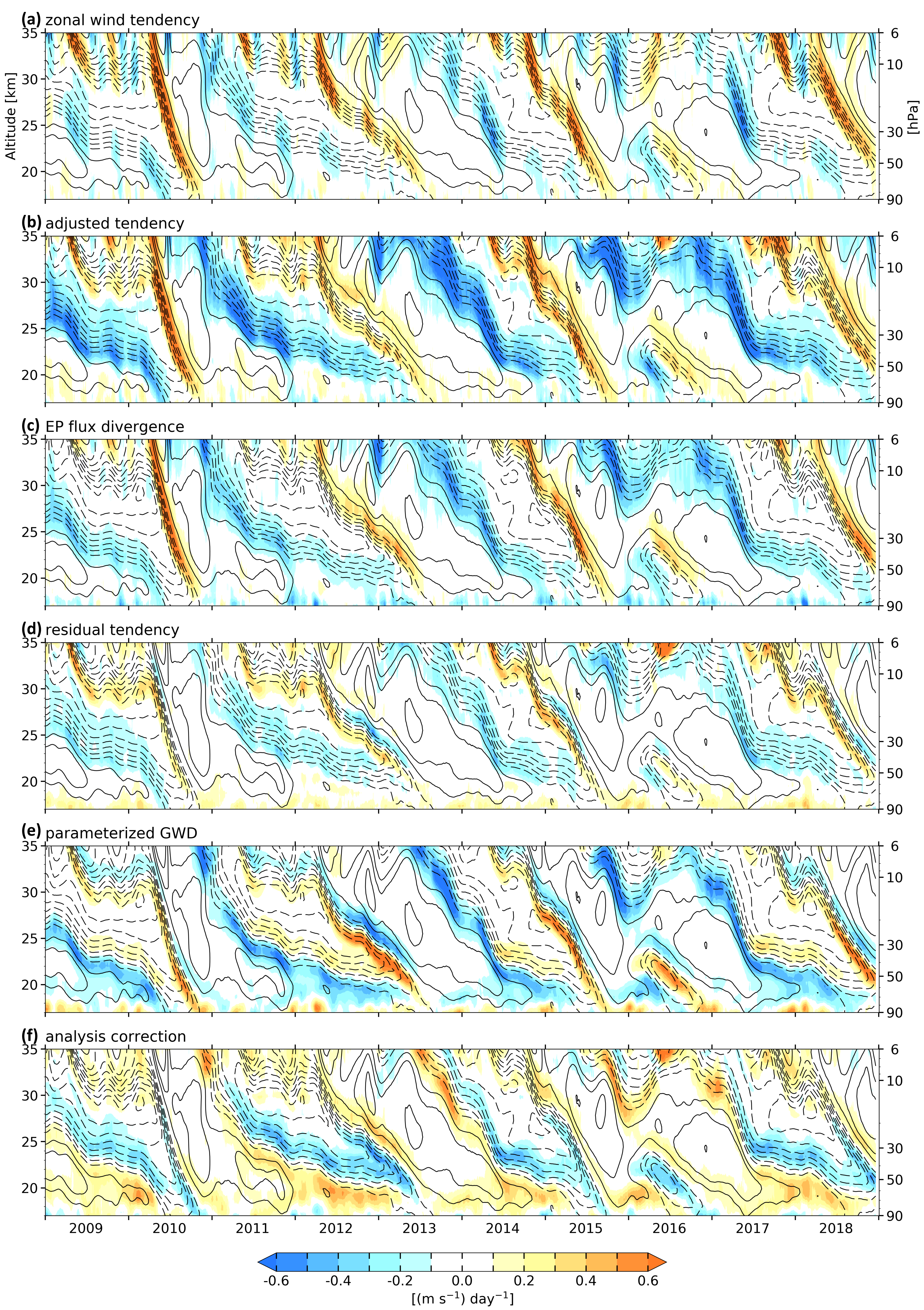}}

\caption{As in Fig. 7 but the zonal wind tendency due to different terms in Eq. (1) indicated by colored shading. See the text for the definition of the adjusted and residual tendencies.} \label{fig9}
\end{figure*}

Figures 9 and 10 present an overview of the momentum budget in the tropical stratosphere. The zonal wind tendency, shown in Figs. 9a and 10a, is arguably the most robust of the terms in the balance because it is large and directly related to the winds assimilated from observations, at least in the lower and middle stratosphere. Values range as high as $\pm$0.7 m s$^{-1}$ d$^{-1}$ for brief periods within strong shear zones. The "adjusted tendency", shown in Figs. 9b and 10b, which is the zonal wind tendency minus total advection, is the tendency that remains to be explained by the wave forcing after the advection by the mean meridional motions is taken-into account. In the easterly shear zones, which are propagating downward in the presence of strong ambient ascent (Fig. 4), the adjusted tendency is appreciably larger than the observed tendency. The wave forcing needs to account for the downward propagation of successive westerly and easterly regimes and it also needs to account for the maintenance of the QBO in the presence of damping (especially of the easterlies) by the advection. How the waves satisfy (or fail to satisfy) these balance requirements, is the focus of the remainder of this section.

\begin{figure*}[t]
\centerline{\includegraphics[width=\textwidth]{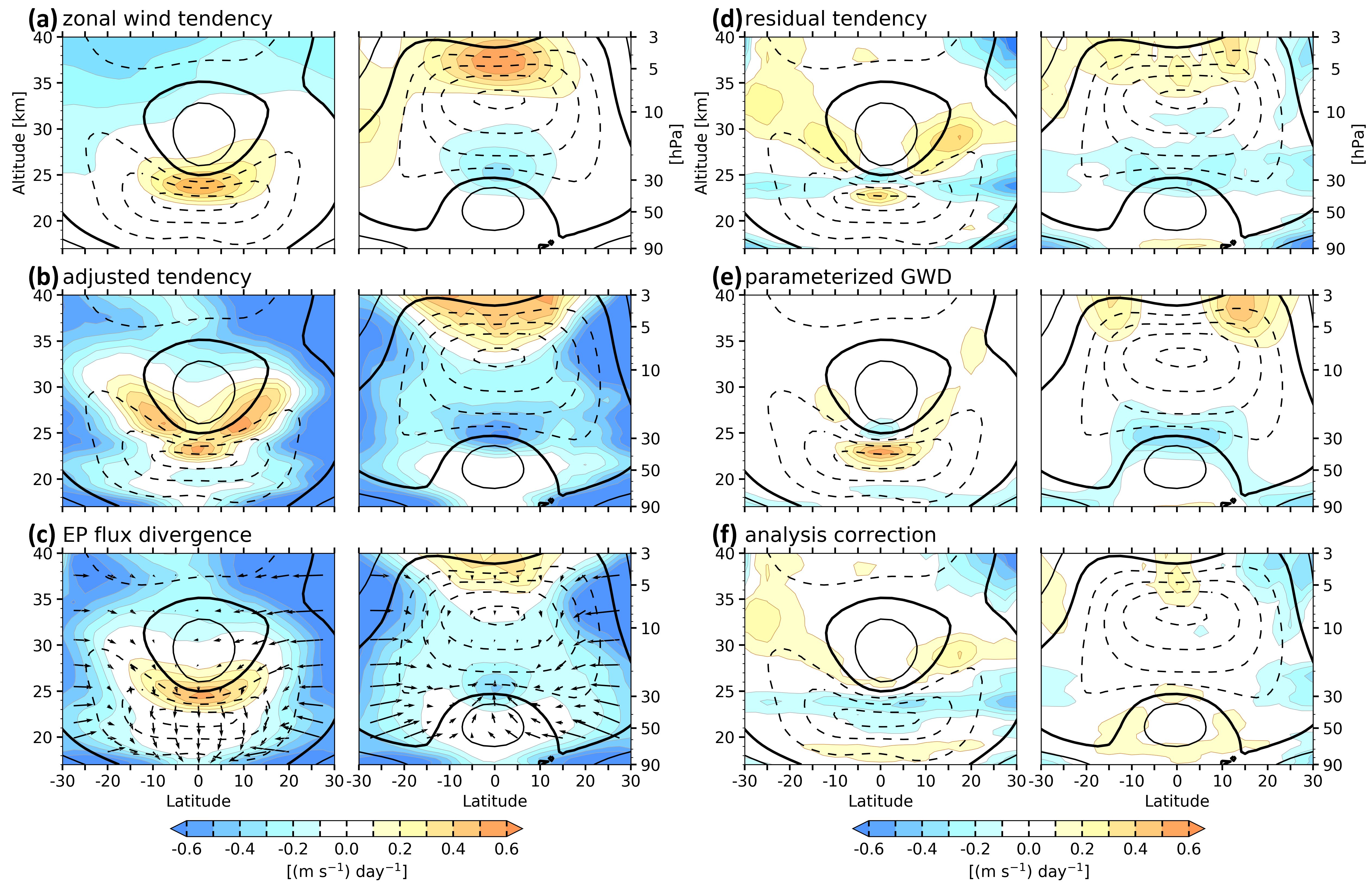}}

\caption{As in Fig. 8 but the zonal wind tendency (colored shading) due to different terms in Eq. (1) as indicated. Arrows in (c) indicate the EP fluxes. In the westerly shear zone (left), the longest vertical components of the arrows correspond to $5.2 \times 10^8$ kg s$^{-2}$, while the longest horizontal components correspond to $1.5 \times 10^{11}$ kg s$^{-2}$. In the easterly shear zone (right), the arrows are scaled to be twice as long. See the text for the definition of the adjusted and residual tendencies.} \label{fig9}
\end{figure*}

\subsection{Forcing by the resolved waves}

The adjusted zonal wind acceleration near the equator due to the wave forcing is due to both resolved and unresolved waves. The resolved forcing is proportional to the divergence of the EP flux, shown in Figs. 9c. The EP flux divergence is concentrated within the shear zones, with values ranging as high as $\pm$0.6 m s$^{-1}$ d$^{-1}$ over intervals of a month or two. These values are generally consistent with results of general circulation models in which waves of short horizontal wavelength are modeled explicitly and no gravity wave parameterization is required (e.g., Kawatani et al. 2010a,b), and they are larger than those based on ERA-I, especially in the easterly shear zones that have been documented in previous studies such as those of Ern et al. (2014) and Kim and Chun (2015). The strong correspondence between the strength of the shear and the strength of the forcing in Fig. 9c is consistent with theoretical expectations: the stronger the vertical shear, the more of the upward-dispersing waves encounter their critical levels per unit height, the more zonal momentum they deposit, and hence, the stronger the zonal acceleration. It is evident from Fig. 9c that the waves also induce a more widespread westward acceleration of a few tenths of a m s$^{-1}$ d$^{-1}$ throughout the lower halves of easterly regimes.

Meridional cross-sections of wave forcing during the phases of the QBO cycle when westerly and easterly shear zones are descending through the 25 hPa level are shown in Fig. 10c. Also shown in these panels is the transport of westerly momentum in the meridional plane, which is in the opposite direction of the EP flux vectors.  In the westerly shear zone composite (left), waves dispersing upward through the easterly regime in the lower stratosphere are absorbed when they encounter the overlying westerly shear zone, inducing a strong eastward acceleration. In a similar manner, in advance of the descending easterly regime (right), westward momentum converges to the shear zone. Weaker convergence is evident throughout most of the easterly regime. In both Figs. 9 and 10, and during the decent of the easterly shear zones, the pattern of resolved wave forcing (c) resembles the adjusted zonal wind tendency (b), but it is not as strong.

\begin{figure*}[t]
\centerline{\includegraphics[width=27pc]{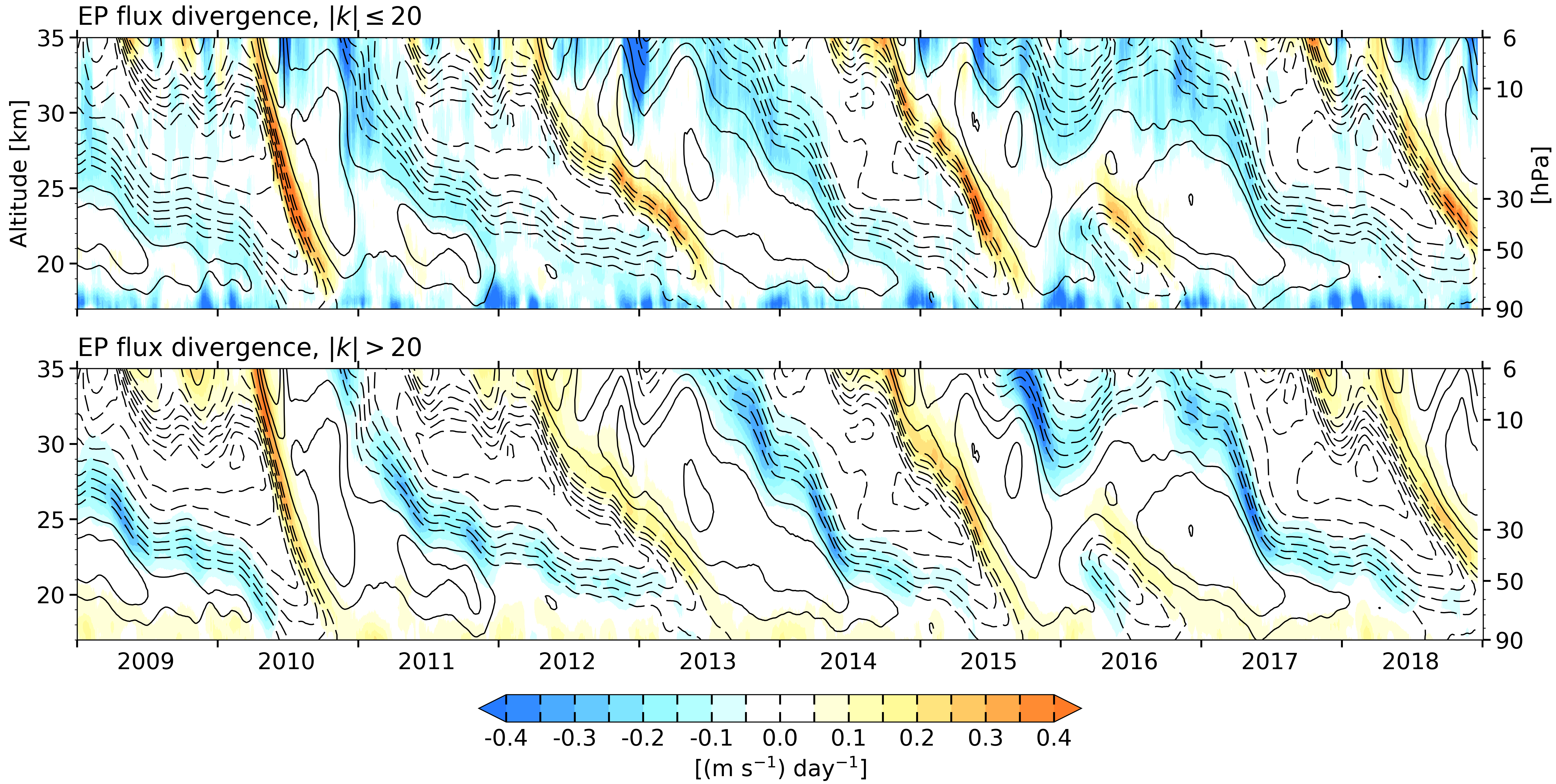}}

\caption{Time-height sections of the EP flux divergence (colored shading) due to planetary-scale waves (top) defined as waves with zonal wavenumbers $|k| \leq 20$, and small-scale gravity (SSG) waves (bottom) defined as waves with zonal wavenumbers $|k| > 20$, averaged over 5\textdegree  N/S. Contours indicate the zonally averaged zonal wind. Contour interval 5 m s$^{-1}$, westerlies solid, easterlies dashed, zero contour omitted. Based on 6-hourly data, smoothed by a 30-day running mean.} \label{fig11}
\end{figure*}

To gain further insight to the relative contribution of waves with different scales to the forcing of the QBO, the resolved wave forcing in Figs. 9c and 10c are separated in Figs. 11 and 12 into contributions from waves with zonal wavenumbers less than and greater than 20, referred to here as "planetary-scale" and "small-scale gravity" (SSG) waves. It is evident that the forcing due to planetary and SSG waves are of comparable importance. The planetary-scale waves make a stronger contribution to the descent of westerly shear zones, while the SSG waves appear to be primarily responsible for the episodes of rapid descent of easterly shear zones, consistent with findings of  Giorgetta et al. (2002).

\begin{figure}[t]
\centerline{\includegraphics[width=19pc]{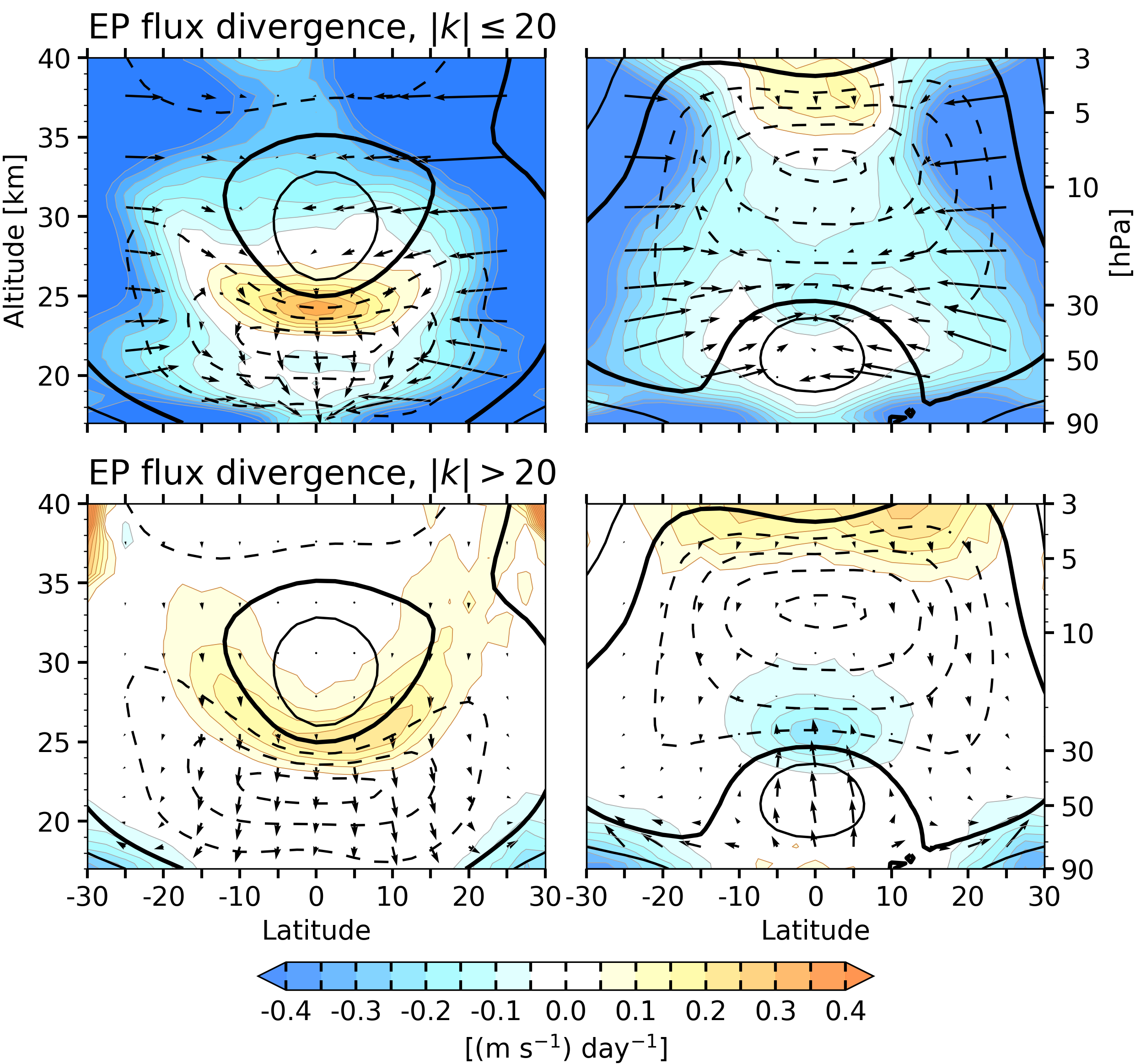}}

\caption{Composites for descending westerly (left) and easterly (right) shear zones of the QBO. The colored shading indicates the EP flux divergence due to planetary-scale waves (top) defined as waves with zonal wavenumbers $|k| \leq 20$, and SSG waves (bottom) defined as waves with zonal wavenumbers $|k| > 20$, superimposed upon the zonally averaged zonal wind (contours). Contour interval 7.5 m s$^{-1}$, westerlies solid, easterlies dashed, zero contour bolded. Arrows indicate the EP fluxes. In the westerly shear zone (left), the longest vertical components of the arrows correspond to $3 \times 10^8$ kg s$^{-2}$, while the longest horizontal components correspond to $1.2 \times 10^{11}$ kg s$^{-2}$. In the easterly shear zone (right), the arrows are scaled to be twice as long.} \label{fig12}
\end{figure}

The patterns of wave forcing in the meridional plane by the two scales of waves shown in Fig. 12 are similar in many respects, but there are some distinctions between them. That the EP flux vectors are quite different suggests that different kinds of equatorially-trapped waves dominate the transports at the different scales, as documented in Part II. While both scales contribute to the eastward accelerations in the descending westerly shear zones, the contribution from the planetary waves is focused in the equatorial belt, whereas the contribution from the SSG waves spans a wider range of latitudes and conforms to the "U" shape of the shear zone. The westward forcing by the planetary-scale waves is broadly distributed within the easterly regimes, whereas the contribution from the SSG waves is more focused in the shear zones, where it plays the dominant role in their descent.

Earlier study also suggests the importance of the SSG waves in the forcing of the QBO. Kawatani et al. (2010a,b) simulated the QBO using a high-resolution atmospheric general circulation model with 60 km horizontal resolution and 300 m vertical resolution, in which waves with short horizontal wavelengths were modeled explicitly so that no gravity wave parameterization was required. They also concluded that the gravity waves with $k$ > 36 are the main contributors to producing the descent of the easterly shear zones in the simulated QBO.

\subsection{Interpretation of the unresolved wave forcing}

The residual zonal wind tendency $\bar X$ in Eq. (1), that remains after the contribution from the resolved waves is subtracted out from the adjusted tendency, is shown in Figs. 9d and 10d (i.e., d =  b - c in Figs. 9 and 10). In studies of Ern et al. (2014), Kim and Chun (2015) and others, the terms used in evaluating $\bar X$ are assumed to have been estimated with sufficient accuracy to warrant interpreting $\bar X$ as the contribution of unresolved gravity waves to the wave forcing. Here we make the same assumption. The similarity between the resolved and the residual forcing during the descent of the easterly shear zones suggests that much of the residual can be explained simply by assuming that the unresolved waves behave in the same manner as the resolved waves. Indeed, by multiplying the resolved wave forcing by a factor of $\sim$1.7, the residual tendency would largely become zero in the easterly shear zones so that the unresolved wave forcing accounts for $\sim$40\% of the total (i.e., resolved plus unresolved) wave forcing in the easterly shear zones. The unresolved waves evidently produce a strong westward forcing in the core of the easterly shear zones, especially during intervals of rapid descent. Results of Giorgetta et al. (2006), Kawatani et al. (2010a), among others, based on numerical models with high spatial resolution also suggest that waves on scales too small to be resolved by ERA5 could play an important role in forcing the descent of easterly regimes.

The residual tendencies are also marked by eastward forcing in a thin layer just above the peaks of the easterly wind regimes, accompanied westward forcing higher up, near the zero-wind line. The existence of this couplet indicates that the unresolved waves must be depositing eastward momentum lower in the westerly shear zones than the resolved waves. Two different, but complementary ideas can be put forth to explain it: (1) that the vertical wind shear in westerly shear zones may become so strong that it exceeds the criterion for Kelvin-Helmholtz (KH) instability, in which case the shear is maintained at the critical value by the downward mixing of eastward momentum in the unresolved waves, (2) that eastward propagating waves dispersing upward through the layer of peak easterlies begin to amplify in inverse proportion to their Doppler-shifted frequency and the waves with the shortest vertical wavelengths are the first to break as the shear within them becomes supercritical.

Other notable features in the pattern of the residual tendency are the patches of eastward forcing along the flanks of the westerly wind regimes (Fig. 10d). Several studies have investigated the possible role of barotropic instability of the westerly jet (e.g., Andrews and McIntyre 1976; Hamilton et al. 2001; Yao and Jablonowski 2015). Indeed, the presence of narrow westerly jets gives rise to a reversal of the zonal mean barotropic vorticity gradient, which is a necessary condition for barotropic instability (Garcia and Richter 2019). The distinct pattern around the westerly jet, with westward forcing near its nose and patches of eastward forcing along its flanks, acts to decelerate the westerly jet close to the equator and to accelerate it away from the equator. Its role would be to reduce the curvature of the westerly jet by broadening its profile, thereby neutralizing the instability inherent in the flow configuration.

The residual wave forcing $\bar X$ can be partitioned into the part that is captured by the gravity wave parameterization scheme and the analysis correction, which can be interpreted as the error in the model's representation of the QBO momentum balance, prior to the data assimilation step in constructing the reanalysis. The parameterized gravity wave drag (GWD) in ERA5 shown in Figs. 9e and 10e captures some of these unresolved features. That there are differences between the patterns in panels (d) and (e) is indicative of systematic (and therefore potentially correctable) biases in the GWD parameterization or in the evaluation of the terms in the balance. Near the equator (Fig. 9e), the GWD parameterization scheme overcompensates for the weakness of the resolved waves. If the GWD parameterization is tuned to fit $\bar X$, the analysis correction would be reduced accordingly. In contrast, at latitudes $\sim$10-20\textdegree  N/S the parametrized GWD is too weak.

\section{The QBO in ERA5 vs. ERA-I }

Here we document the results of a detailed comparison of the representations of the QBO in ERA5 and in ERA-I. The difference in monthly mean zonally averaged zonal wind, averaged over 5\textdegree  N/S, is shown in Fig. S2. Prior to 1998, the SAO-related winds are systematically stronger in ERA5, particularly during the westerly phase. The QBO-related winds are also stronger in ERA5 prior to 1998, particularly near the shear zones in the easterly regimes. The differences decrease with time during the 1999-2006 period, and a good agreement can be seen thereafter. It should be borne in mind that throughout this period of record, fewer data are assimilated in reanalysis datasets at the higher altitudes because it is difficult to measure winds at these heights based on satellite or ground-based remote sensing (Anstey et al. 2020). Satellite direct wind measurements have not been available above $\sim$30 km (Smith et al. 2017) and nadir-viewing satellite observations have difficulty resolving the temperature perturbations in the layers of strong vertical wind shear in the SAO, creating large uncertainties in the SAO winds (Coy et al. 2016).

\begin{figure*}[t]
\centerline{\includegraphics[width=27pc]{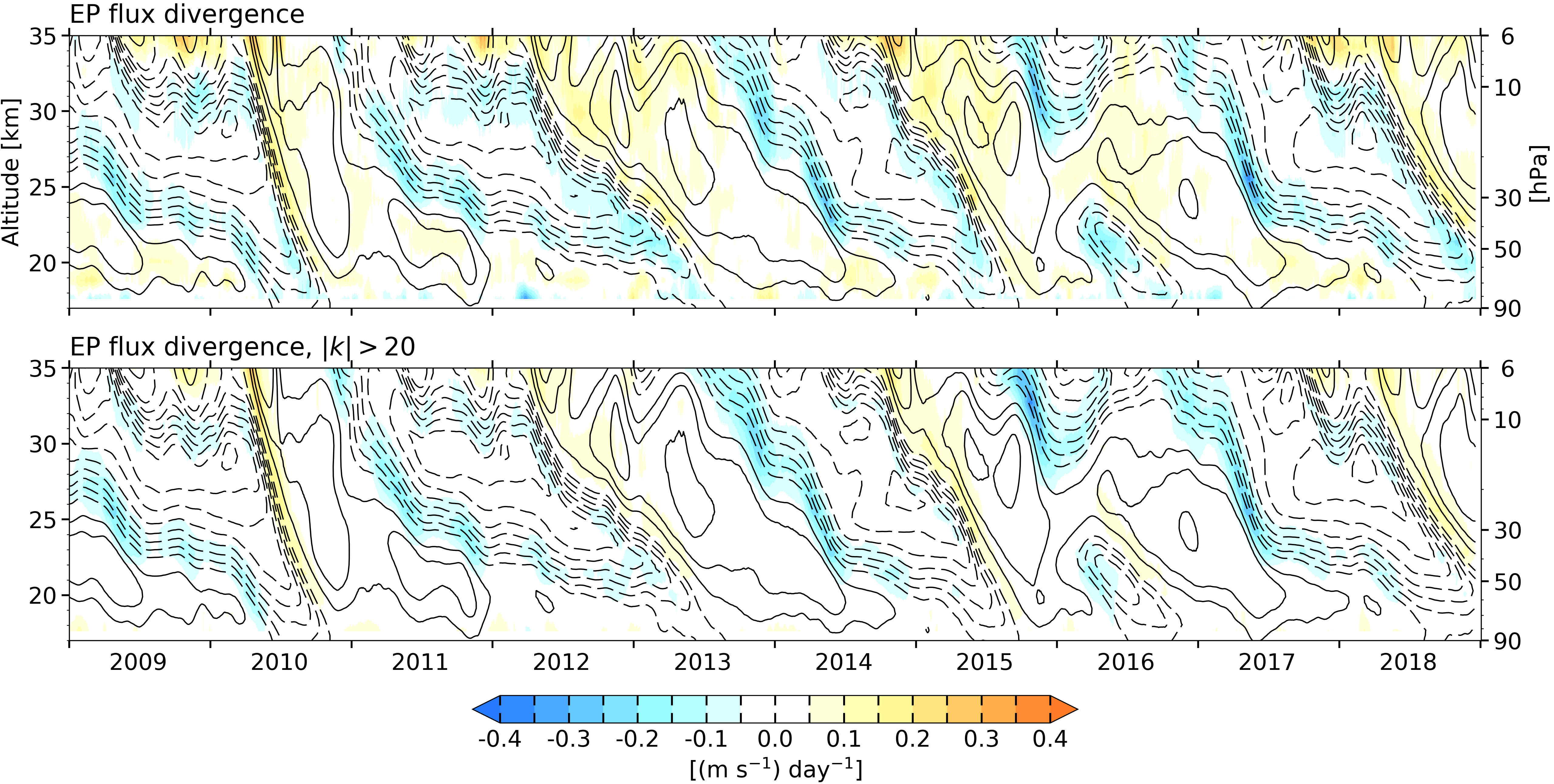}}

\caption{Time-height sections of the difference between ERA5 and ERA-I in EP flux divergence (colored shading) due to all resolved waves (top), and SSG waves (bottom) defined as waves with zonal wavenumbers $|k| > 20$, averaged over 5\textdegree  N/S. Contours indicate the zonally averaged zonal wind. Contour interval 5 m s$^{-1}$, westerlies solid, easterlies dashed, zero contour omitted. Based on 6-hourly data, smoothed by a 30-day running mean.} \label{fig13}
\end{figure*}

\begin{figure}[t]
\centerline{\includegraphics[width=19pc]{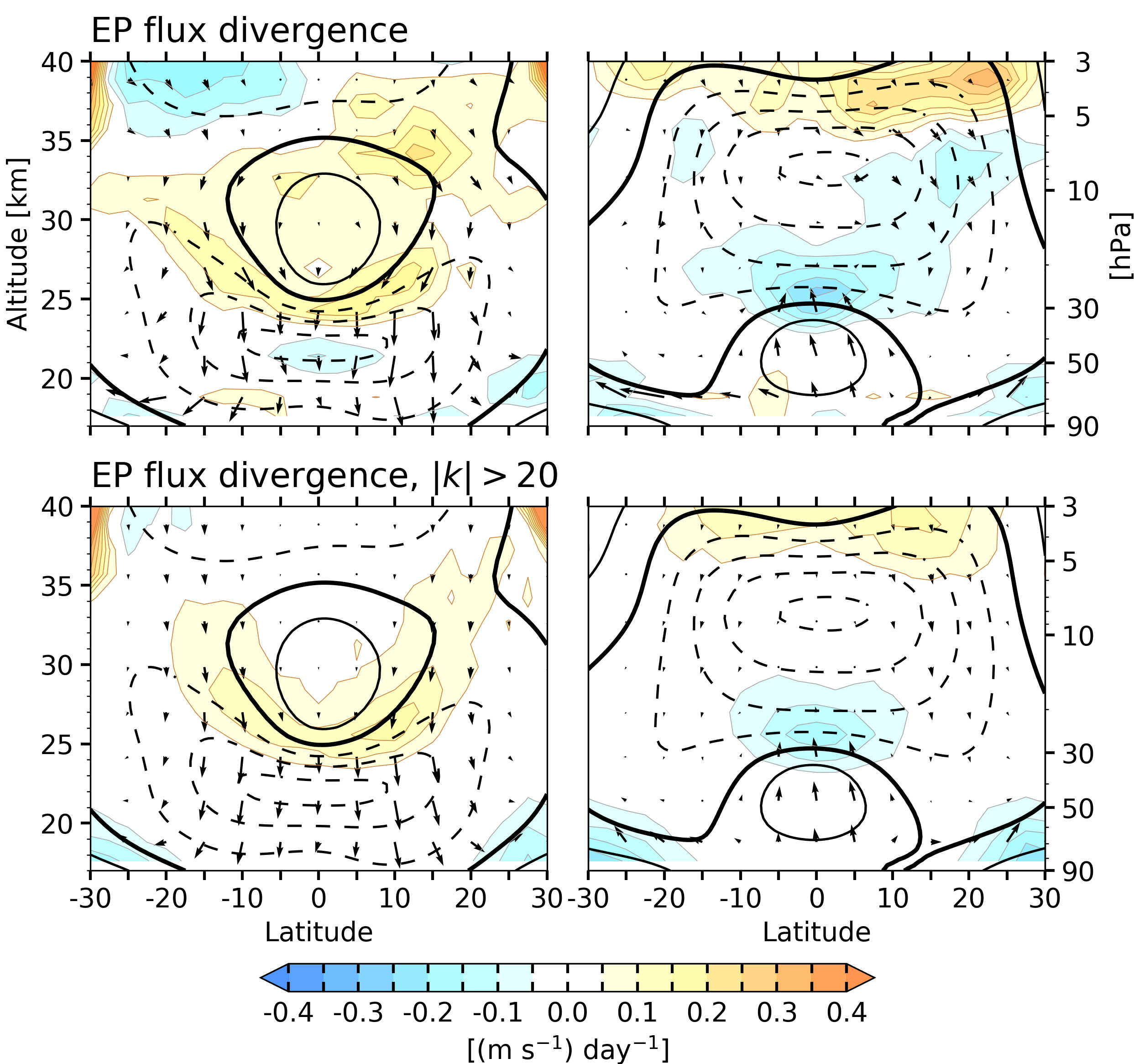}}

\caption{Composites for descending westerly (left) and easterly (right) shear zones of the QBO. The colored shading indicates the difference between ERA5 and ERA-I in EP flux divergence due to all resolved waves (top), and SSG waves (bottom) defined as waves with zonal wavenumbers $|k| > 20$, superimposed upon the zonally averaged zonal wind (contours). Contour interval 7.5 m s$^{-1}$, westerlies solid, easterlies dashed, zero contour bolded. Arrows indicate the EP fluxes. In the westerly shear zone (left), the longest vertical components of the arrows correspond to $2.2 \times 10^8$ kg s$^{-2}$, while the longest horizontal components correspond to $3.2 \times 10^{10}$ kg s$^{-2}$.} \label{fig14}
\end{figure}

The analyses presented for ERA5 in the text were also performed on ERA-I and the results are presented in the supplemental material along with their ERA5 counterparts to make the comparison easier. Overall, the differences in the mean fields are subtle, as documented in Figs. S3 to S6. The patterns of adjusted zonal wind tendency and resolved wave forcing in ERA-I and ERA5 are similar in many respects (Figs. S7 to S9). However, the resolved wave forcing in ERA5 is stronger, and accounts for a substantially larger fraction of the adjusted zonal wind tendency.
A more quantitative comparison between the resolved wave forcing in the two reanalysis is presented in Figs. 13 and 14 in which EP flux divergence in ERA-I is subtracted from that in ERA5. For reference, ERA5 is capable of resolving waves with wavenumbers ranging up to 360, compared to 120 in ERA-I. It is evident that forcing in ERA5 is stronger by up to 0.3 m s$^{-1}$ d$^{-1}$ during the descent of the easterly shear zones and up to 0.2 m s$^{-1}$ d$^{-1}$ during the descent of the westerly shear zones. Generally speaking, the forcing in ERA5 more faithfully captures the spikes in wave forcing that gives rise to short-lived episodes in which easterly regimes descend as rapidly as westerly regimes. The bottom panels reveal that this difference is mainly attributable to forcing from SSG waves.

Indeed, we find the forcing due to SSG waves in ERA-I to be much reduced (Figs. S10 and S11), in particular in the easterly shear zones, consistent with the results of Ern et al. (2014), who calculated the EP flux divergence by waves $k$ = 1-20 in ERA-I and found the contribution from the wavenumbers higher than that to be relatively unimportant. The planetary-scale forcing is also slightly smaller in ERA-I than in ERA5 (Figs. S9 and S10), which is probably due to lower vertical resolution of ERA-I, resulting in a poorer representation of wave-mean flow interaction processes.

\section{Discussion and conclusion }

This study has documented several aspects of the QBO and provided deeper insights into its momentum budget. It is shown that the QBO-induced MMC, while mainly confined to the winter hemisphere, is strong enough to modify the BDC-related upwelling. Within 5\textdegree  of the equator the descent in the westerly shear zones interrupts the tropical upwelling and deflects the flow around the flanks of the westerly regimes. In the easterly shear zones the tropical upwelling is enhanced. The QBO-induced MMC creates and sustains the QBO-related temperature perturbations against radiative damping, while it damps the QBO-related zonal wind perturbations. Consequently, the wave forcing is responsible for both the downward propagation and the maintenance of the QBO.

The resolved waves in ERA5 produce roughly comparable contributions to the eastward and westward accelerations, which range as high as  $\pm$0.6 m s$^{-1}$ d$^{-1}$ during the descent of the westerly and easterly shear zones of the QBO. Separating ERA5 resolved wave forcing into contributions from SSG waves with zonal wavelengths less than 2000 km and planetary-scale waves with wavelengths longer than that reveals that the planetary scale waves make a stronger contribution to the descent of westerly shear zones, while SSG waves play the dominant role in the descent of easterly shear zones.

The time-height section for the analysis correction shown in Fig. 9f exhibits a distinctive pattern during the extended intervals in which a westerly regime is temporarily stalled at the 20 km level while the core of the easterly regime that will eventually replace it is stalled in the 27-30 km layer. Under these conditions the model used in generating the ERA5 reanalyses produces a spurious westward acceleration in the lingering westerly regime and an eastward acceleration in the stalled easterly shear zone centered  $\sim$23 km (Fig. 9e), both of which need to be corrected by the analysis increment. It is evident from Fig. 9f that much of the required analysis correction involves a cancellation of the spurious forcing imposed by the GWD parameterization scheme.

\begin{table*}[t]
\caption{Variance of tendencies [$10^{20}$ (m$^2$ s$^{-2}$) day$^{-2}$], averaged over 5\textdegree  N/S, integrated over time (1979 to 2018) and height (10-90 hPa), and their ratio with respect to the adjusted tendency (parentheses). See the text for explanation.}
\begin{tabular*}{\textwidth}{@{\extracolsep\fill}lcccccc@{}}
\topline
& zonal wind tendency& adjusted tendency &EP flux divergence& residual tendency&parametrized GWD& analysis correction\\
\midline
\ ERA5& 5.1& 11.3& 5.6 (50\%)& 3.7 (33\%)& 6.4& 4.4\\
\ ERA-I& 4.9& 10.5& 2.7 (26\%)& 4.9 (47\%)& - &-\\
\botline
\end{tabular*}
\end{table*}

\begin{figure}[t]
\centerline{\includegraphics[width=10pc]{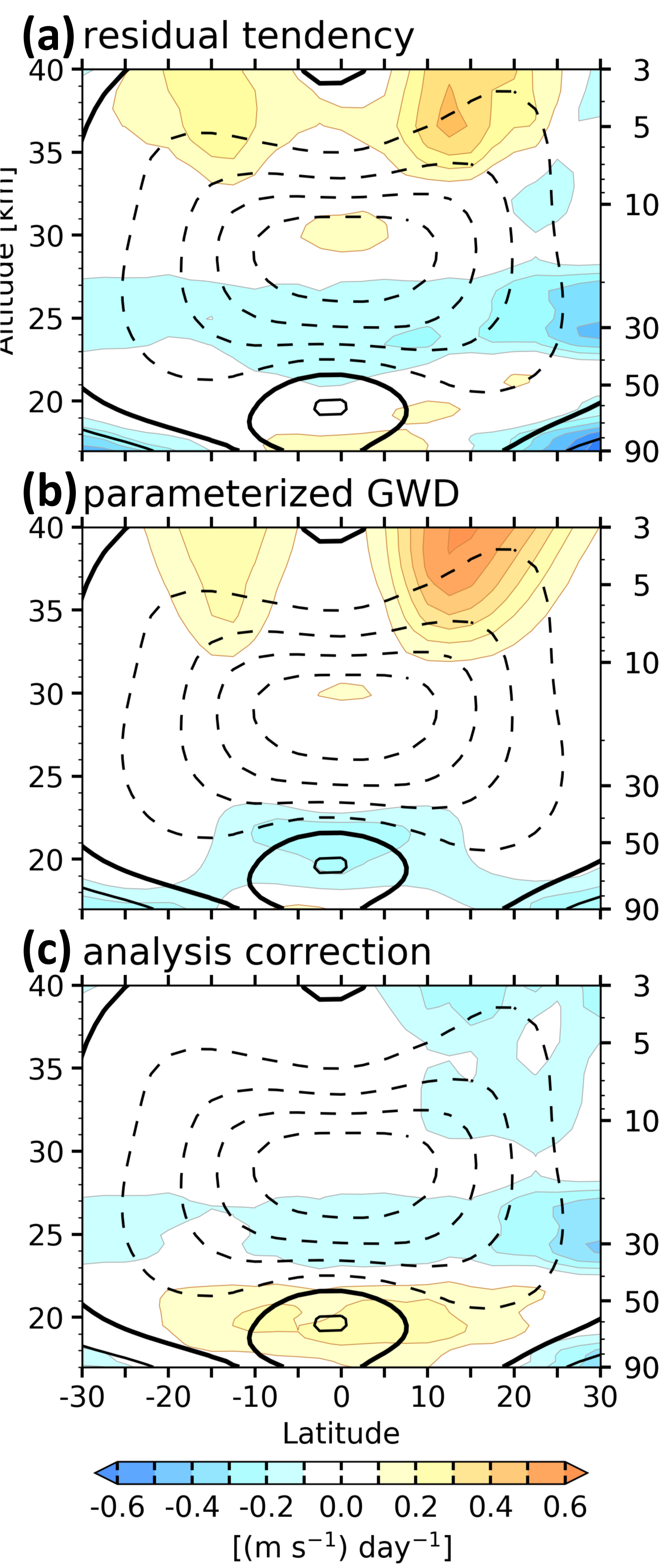}}

\caption{Composites for descending easterly shear zone of the QBO, as defined in the text. The colored shading shows the zonal wind tendency due to the indicated terms, superimposed upon the zonally averaged zonal wind (contours). Contour interval 7.5 m s$^{-1}$, westerlies solid, easterlies dashed, zero contour bolded.} \label{fig12}
\end{figure}

This problem is not apparent in the meridional cross section for the easterly shear zone shown in Fig. 10f because the composite upon which it is based includes only rapidly descending shear zones (see Fig. S1). To show the off-equatorial structure during these intervals, we defined a new composite based on the 120 months of the 480-month period of record when the easterly jet at 27-30 km level was strongest (Fig. S12). The results shown in Fig. 15 are generally consistent with the time-height sections in Fig. 9, including the analysis correction. The features in the equatorial time-height sections extend out to 15-20\textdegree  N/S.

As previously discussed, the terms used in evaluating the residual tendency $\bar X$ (see Eq. (1)), are assumed to have been estimated with sufficient accuracy to justify interpreting the residual tendency as the contribution of unresolved gravity waves to the wave forcing. While It is encouraging to see some similarity in the parameterized GWD and residual tendency patterns (Figs. 9d,e, Figs. 10d,e, Figs. 15a,b), it is evident from the analysis correction that significant discrepancies remain. For example, the westward forcing in the parameterized GWD in the easterly shear zone is too strong and too low, leading to a dipole of westward and eastward forcing in the lower stratosphere in the analysis correction (see Fig.15c and Fig. 9f).

The representation of the mean fields in the QBO is found to be very similar in ERA5 and ERA-I, but ERA5\textquotesingle s higher spatial resolution makes it possible to resolve a broader spectrum of atmospheric waves. With its higher vertical resolution, it is also possible to represent the process of wave-mean flow interaction more realistically. As shown in Table 1, the resolved waves in ERA5 (i.e., the EP flux divergence) explains half of the variance of the adjusted tendency, indicating a significant improvement over ERA-I. The variance attributable to unresolved waves (i.e., the residual tendency) has decreased in ERA5 accordingly. That the sum of the variances of the EP flux divergence and the residual tendency is smaller than the variance of the adjusted tendency is indicative of their nonzero covariance in the time-height domain. This is to be expected, since the pattern of the residual tendency projects positively upon the pattern of the EP flux divergence, as shown in Fig. 9. If the covariance were eliminated by artificially inflating the amplitude of the EP flux divergence by a factor of 1.18, the fraction of the variance "explained" by the resolved waves would increase from 50\% to 69\% and the residual "unexplained" variance would drop to 31\%. However, locally large analysis corrections would still be required in the westerly shear zones. The GWD parameterization scheme accounts for much of the unresolved forcing in the westerly shear zones, but in the present version it produces an excessively strong westward forcing in the rapidly descending easterly shear zones, and it prematurely damps the stalled westerly regimes at 50 hPa and below. To further reduce the analysis correction, it will thus be necessary both to further increase the resolution of the model, especially in the vertical, and to further refine the GWD parameterizations scheme.

\acknowledgments
The authors would like to thank M. Joan Alexander, Peter H. Haynes, and Pu Lin for the helpful discussions and suggestions. This research is supported by the NASA grant 80NSSC18K1031 and NSF grant AGS-1821437. The authors declare no conflicts of interest.

\datastatement
ERA5 data were downloaded from ECMWF's MARS archive. Aura MLS observations were obtained from NASA Goddard Earth Sciences (GES) Data and Information Services Center (DISC). CALIPSO data were downloaded using NASA Search and Subsetting Web Application (https://subset.larc.nasa.gov/calipso/).

\bigskip

 \textbf{References}

\bigskip

Andrews, D. G., and M. F. McIntyre, 1976: Planetary waves in horizontal and vertical shear: Asymptotic theory for equatorial waves in weak shear. Journal of the Atmospheric Sciences, 33, 2049-2053.

------, J. D. Mahlman, and R. W. Sinclair, 1983: Eliassen-Palm Diagnostics of Wave-Mean Flow Interaction in the GFDL SKYHI General Circulation Model. J. Atmos. Sci., 40, 2768-2784.

Andrews, D. G., C. B. Leovy, and J. R. Holton, 1987: Middle atmosphere dynamics. Academic press.

Angell, J. K., and J. Korshover, 1964: Quasi-biennial variations in temperature, total ozone, and tropopause height. Journal of the Atmospheric Sciences, 21, 479-492.

Baldwin, M. P., and Coauthors, 2001: The quasi-biennial oscillation. Reviews of Geophysics, 39, 179-229.

Booker, J. R., and F. P. Bretherton, 1967: The critical layer for internal gravity waves in a shear flow. Journal of Fluid Mechanics, 27, 513-539.

Brewer, A. W., 1949: Evidence for a world circulation provided by the measurements of helium and water vapour distribution in the stratosphere. Quarterly Journal of the Royal Meteorological Society, 75, 351-363, https://doi.org/10.1002/qj.49707532603.

Bushell, A. C., and Coauthors, 2020: Evaluation of the Quasi-Biennial Oscillation in global climate models for the SPARC QBO-initiative. Quarterly Journal of the Royal Meteorological Society,.

Coy, L., K. Wargan, A. M. Molod, W. R. McCarty, and S. Pawson, 2016: Structure and dynamics of the quasi-biennial oscillation in MERRA-2. Journal of climate, 29, 5339-5354.

Delisi, D. P., and T. J. Dunkerton, 1988: Seasonal Variation of the Semiannual Oscillation. J. Atmos. Sci., 45, 2772-2787.

Dunkerton, T. J., 1997: The role of gravity waves in the quasi-biennial oscillation. Journal of Geophysical Research: Atmospheres, 102, 26053-26076.

-----, 2016: The quasi-biennial oscillation of 2015-2016: Hiccup or death spiral? Geophysical Research Letters, 43, 10-547.

-----, and D. P. Delisi, 1997: Interaction of the quasi-biennial oscillation and stratopause semiannual oscillation. Journal of Geophysical Research: Atmospheres, 102, 26107-26116.

Ern, M., and P. Preusse, 2009a: Quantification of the contribution of equatorial Kelvin waves to the QBO wind reversal in the stratosphere. Geophysical research letters, 36.

-----, and -----, 2009b: Wave fluxes of equatorial Kelvin waves and QBO zonal wind forcing derived from SABER and ECMWF temperature space-time spectra. Atmos. Chem. Phys, 9, 3957-3986.

-----, and Coauthors, 2014: Interaction of gravity waves with the QBO: A satellite perspective. Journal of Geophysical Research: Atmospheres, 119, 2329-2355.

Fleming, E. L., C. H. Jackman, J. E. Rosenfield, and D. B. Considine, 2002: Two-dimensional model simulations of the QBO in ozone and tracers in the tropical stratosphere. Journal of Geophysical Research: Atmospheres, 107, ACL-1.

Fueglistaler, S., and P. H. Haynes, 2005: Control of interannual and longer-term variability of stratospheric water vapor. Journal of Geophysical Research: Atmospheres, 110.

Garcia, R. R., and J. H. Richter, 2019: On the momentum budget of the quasi-biennial oscillation in the whole atmosphere community climate model. Journal of the Atmospheric Sciences, 76, 69-87.

------, T. J. Dunkerton, R. S. Lieberman, and R. A. Vincent, 1997: Climatology of the semiannual oscillation of the tropical middle atmosphere. Journal of Geophysical Research: Atmospheres, 102, 26019-26032.

Giorgetta, M. A., E. Manzini, and E. Roeckner, 2002: Forcing of the quasi-biennial oscillation from a broad spectrum of atmospheric waves. Geophysical Research Letters, 29, 86-1.

Giorgetta, M. A., E. Manzini, E. Roeckner, M. Esch, and L. Bengtsson, 2006: Climatology and forcing of the quasi-biennial oscillation in the MAECHAM5 model. Journal of Climate, 19, 3882-3901.

Hamilton, K., 1984: Mean wind evolution through the quasi-biennial cycle in the tropical lower stratosphere. Journal of the atmospheric sciences, 41, 2113-2125.

-----, R. J. Wilson, and R. S. Hemler, 2001: Spontaneous stratospheric QBO-like oscillations simulated by the GFDL SKYHI general circulation model. Journal of the atmospheric sciences, 58, 3271-3292.

Hampson, J., and P. Haynes, 2004: Phase Alignment of the Tropical Stratospheric QBO in the Annual Cycle. J. Atmos. Sci., 61, 2627-2637, https://doi.org/10.1175/JAS3276.1.

Hartmann, D. L., L. A. Moy, and Q. Fu, 2001: Tropical Convection and the Energy Balance at the Top of the Atmosphere. J. Climate, 14, 4495-4511, https://doi.org/10.1175/1520-0442(2001)014<4495:TCATEB>2.0.CO;2.

Hersbach, H., and Coauthors, 2020: The ERA5 global reanalysis. Quarterly Journal of the Royal Meteorological Society, 146, 1999-2049, https://doi.org/10.1002/qj.3803.

Hitchman, M. H., and A. S. Huesmann, 2007: A Seasonal Climatology of Rossby Wave Breaking in the 320-2000-K Layer. J. Atmos. Sci., 64, 1922-1940, https://doi.org/10.1175/JAS3927.1.

Hoffmann, L., and Coauthors, 2019: From ERA-Interim to ERA5: the considerable impact of ECMWF’s next-generation reanalysis on Lagrangian transport simulations. Atmospheric Chemistry and Physics, 19, 3097-3124.

Holt, L. A., M. J. Alexander, L. Coy, A. Molod, W. Putman, and S. Pawson, 2016: Tropical waves and the quasi-biennial oscillation in a 7-km global climate simulation. Journal of the Atmospheric Sciences, 73, 3771-3783.

-----, and Coauthors, 2020: An evaluation of tropical waves and wave forcing of the QBO in the QBOi models. Quarterly Journal of the Royal Meteorological Society, n/a, https://doi.org/10.1002/qj.3827.

Holton, J. R., 1989: Influence of the annual cycle in meridional transport on the quasi-biennial oscillation in total ozone. Journal of the Atmospheric Sciences, 46, 1434-1439.

-----, and R. S. Lindzen, 1972: An updated theory for the quasi-biennial cycle of the tropical stratosphere. Journal of the Atmospheric Sciences, 29, 1076-1080.

-----, and W. M. Wehrbein, 1980: A numerical model of the zonal mean circulation of the middle atmosphere. pure and applied geophysics, 118, 284-306.

Kawatani, Y., S. Watanabe, K. Sato, T. J. Dunkerton, S. Miyahara, and M. Takahashi, 2010a: The roles of equatorial trapped waves and internal inertia-gravity waves in driving the quasi-biennial oscillation. Part II: Three-dimensional distribution of wave forcing. Journal of the atmospheric sciences, 67, 981-997.

-------, -------, -------, ------, ------, and -----, 2010b: The roles of equatorial trapped waves and internal inertia-gravity waves in driving the quasi-biennial oscillation. Part I: Zonal mean wave forcing. Journal of the atmospheric sciences, 67, 963-980.

Kim, Y.-H., and H.-Y. Chun, 2015: Momentum forcing of the quasi-biennial oscillation by equatorial waves in recent reanalyses. Atmospheric Chemistry and Physics, 15.

Kinnersley, J. S., and S. Pawson, 1996: The descent rates of the shear zones of the equatorial QBO. Journal of the atmospheric sciences, 53, 1937-1949.

Krismer, T. R., M. A. Giorgetta, and M. Esch, 2013: Seasonal aspects of the quasi-biennial oscillation in the Max Planck Institute Earth System Model and ERA-40. Journal of Advances in Modeling Earth Systems, 5, 406-421.

Lin, L., Q. Fu, H. Zhang, J. Su, Q. Yang, and Z. Sun, 2013: Upward mass fluxes in tropical upper troposphere and lower stratosphere derived from radiative transfer calculations. Journal of Quantitative Spectroscopy and Radiative Transfer, 117, 114-122, https://doi.org/10.1016/j.jqsrt.2012.11.016.

Lindzen, R. S., and J. R. Holton, 1968: A theory of the quasi-biennial oscillation. Journal of the Atmospheric Sciences, 25, 1095-1107.

Maruyama, T., 1991: Annual and QBO-synchronized variations of lower-stratospheric equatorial wave activity over Singapore during 1961-1989. Journal of the Meteorological Society of Japan. Ser. II, 69, 219-232.

Match, A., and S. Fueglistaler, 2019: The Buffer Zone of the Quasi-Biennial Oscillation. J. Atmos. Sci., 76, 3553-3567, https://doi.org/10.1175/JAS-D-19-0151.1.

Mayr, H. G., J. G. Mengel, K. L. Chan, and F. T. Huang, 2010: Middle atmosphere dynamics with gravity wave interactions in the numerical spectral model: Zonal-mean variations. Journal of atmospheric and solar-terrestrial physics, 72, 807-828.

Molod, A., L. Takacs, M. Suarez, and J. Bacmeister, 2015: Development of the GEOS-5 atmospheric general circulation model: evolution from MERRA to MERRA2. Geoscientific Model Development, 8, 1339-1356, https://doi.org/10.5194/gmd-8-1339-2015.

Mote, P. W., and Coauthors, 1996: An atmospheric tape recorder: The imprint of tropical tropopause temperatures on stratospheric water vapor. Journal of Geophysical Research: Atmospheres, 101, 3989-4006.

Nie, J., and A. H. Sobel, 2015: Responses of Tropical Deep Convection to the QBO: Cloud-Resolving Simulations. J. Atmos. Sci., 72, 3625-3638, https://doi.org/10.1175/JAS-D-15-0035.1.

Park, M., and Coauthors, 2017: Variability of stratospheric reactive nitrogen and ozone related to the QBO. Journal of Geophysical Research: Atmospheres, 122, 10-103.

Pascoe, C. L., L. J. Gray, S. A. Crooks, M. N. Juckes, and M. P. Baldwin, 2005: The quasi-biennial oscillation: Analysis using ERA-40 data. Journal of Geophysical Research: Atmospheres, 110.

Pawson, S., and M. Fiorino, 1998: A comparison of reanalyses in the tropical stratosphere. Part 2: The quasi-biennial oscillation. Climate dynamics, 14, 645-658.

Plumb, R. A., 1977: The interaction of two internal waves with the mean flow: Implications for the theory of the quasi-biennial oscillation. Journal of the Atmospheric Sciences, 34, 1847-1858.

-----, and A. D. McEwan, 1978: The instability of a forced standing wave in a viscous stratified fluid: A laboratory analogue of. the quasi-biennial oscillation. Journal of the atmospheric sciences, 35, 1827-1839.

Randel, W. J., and F. Wu, 1996: Isolation of the ozone QBO in SAGE II data by singular-value decomposition. Journal of the atmospheric sciences, 53, 2546-2559.

------, -----, J. M. Russell III, A. Roche, and J. W. Waters, 1998: Seasonal cycles and QBO variations in stratospheric CH4 and H2O observed in UARS HALOE data. Journal of the Atmospheric Sciences, 55, 163-185.

——, ——, R. Swinbank, J. Nash, and A. O’Neill, 1999: Global QBO circulation derived from UKMO stratospheric analyses. Journal of the atmospheric sciences, 56, 457-474.

——, ——, and D. J. Gaffen, 2000: Interannual variability of the tropical tropopause derived from radiosonde data and NCEP reanalyses. Journal of Geophysical Research: Atmospheres, 105, 15509-15523.

——, R. R. Garcia, and F. Wu, 2002: Time-Dependent Upwelling in the Tropical Lower Stratosphere Estimated from the Zonal-Mean Momentum Budget. J. Atmos. Sci., 59, 2141-2152, https://doi.org/10.1175/1520-0469(2002)059<2141:TDUITT>2.0.CO;2.

Reed, R. J., 1962: Some features of the annual temperature regime in the tropical stratosphere. Mon. Weather Rev, 90, 211-215.

Reed, R. J., 1964: A tentative model of the 26-month oscillation in tropical latitudes. Quarterly Journal of the Royal Meteorological Society, 90, 441-466.

Reed, R. J., 1966: Zonal wind behavior in the equatorial stratosphere and lower mesosphere. Journal of Geophysical Research, 71, 4223-4233.

Reid, G. C., 1994: Seasonal and interannual temperature variations in the tropical stratosphere. Journal of Geophysical Research: Atmospheres, 99, 18923-18932, https://doi.org/10.1029/94JD01830.

Ribera, P., C. Peña-Ortiz, R. Garcia-Herrera, D. Gallego, L. Gimeno, and E. Hernández, 2004: Detection of the secondary meridional circulation associated with the quasi-biennial oscillation. Journal of Geophysical Research: Atmospheres, 109.

Schenzinger, V., S. Osprey, L. Gray, and N. Butchart, 2017: Defining metrics of the Quasi-Biennial Oscillation in global climate models. Geoscientific Model Development, 10.

Schoeberl, M. R., and Coauthors, 2008: QBO and annual cycle variations in tropical lower stratosphere trace gases from HALOE and Aura MLS observations. Journal of Geophysical Research: Atmospheres, 113.

Skamarock, W. C., S.-H. Park, J. B. Klemp, and C. Snyder, 2014: Atmospheric kinetic energy spectra from global high-resolution nonhydrostatic simulations. Journal of the Atmospheric Sciences, 71, 4369-4381.

Smith, A. K., R. R. Garcia, A. C. Moss, and N. J. Mitchell, 2017: The semiannual oscillation of the tropical zonal wind in the middle atmosphere derived from satellite geopotential height retrievals. Journal of the Atmospheric Sciences, 74, 2413-2425.

Swinbank, R., and A. O’Neill, 1994: A stratosphere-troposphere data assimilation system. Monthly Weather Review, 122, 686-702.

Tseng, H.-H., and Q. Fu, 2017: Temperature control of the variability of tropical tropopause layer cirrus clouds. Journal of Geophysical Research: Atmospheres, 122, 11-062.

Wallace, J. M., 1967: A note on the role of radiation in the biennial oscillation. Journal of the Atmospheric Sciences, 24, 598-599.

——, 1973: General circulation of the tropical lower stratosphere. Reviews of Geophysics, 11, 191-222.

Yao, W., and C. Jablonowski, 2015: Idealized quasi-biennial oscillations in an ensemble of dry GCM dynamical cores. Journal of the Atmospheric Sciences, 72, 2201-2226.

\end{document}